\documentclass[letterpaper, 10 pt, conference]{IEEEconf}  

\IEEEoverridecommandlockouts                              
\usepackage[OT1]{fontenc} 
\usepackage{graphics} 
\usepackage{amsmath} 
\usepackage{amssymb}  
\usepackage{tikz,pgfplots}
\usetikzlibrary{calc}
\usepackage{subcaption}
\usepackage{centernot}
\usepackage{cite}
\usepackage{soul}
\usetikzlibrary{positioning}
\usetikzlibrary{patterns}
\usetikzlibrary{automata, positioning, arrows}
\usepackage{cleveref}
\usepackage{algorithm}
\usepackage{mdframed}
\newmdenv[
linecolor=orange,
linewidth=5,
topline=false,
bottomline=false,
leftline=false,
leftmargin=-10pt,
rightmargin=-13pt]{sideruler}
\newmdenv[
linecolor=orange,
linewidth=5,
topline=false,
bottomline=false,
rightline=false,
leftmargin=-13pt,
rightmargin=-10pt]{siderulel}

\usepackage[columnwise]{lineno}

\newtheorem{hyp}{Hypothesis}

\let\labelindent\relax
\usepackage{enumitem}

\newcommand\copyrighttext{%
	\footnotesize \textcopyright2018 IEEE:
	Personal use of this material is permitted. Permission from IEEE must be obtained for all other uses, in
	any current or future media, including reprinting/republishing this material for advertising or promotional
	purposes, creating new collective works, for resale or redistribution to servers or lists, or reuse of any
	copyrighted component of this work in other works.}
\newcommand\copyrightnotice{%
	\begin{tikzpicture}[remember picture,overlay]
		\node[anchor=south,yshift=10pt] at (current page.south) {\fbox{\parbox{\dimexpr\textwidth-\fboxsep-\fboxrule\relax}{\copyrighttext}}};
	\end{tikzpicture}%
}

\title{\LARGE \bf
Multi-objective path tracking control for car-like vehicles with 
differentially bounded $n$-smooth output
}

\author{Karin Festl\authorrefmark{1}, Michael Stolz\authorrefmark{1}\authorrefmark{2} and Daniel Watzenig\authorrefmark{1} \IEEEmembership{Senior~Member,~IEEE}
\thanks{This paper is part of the AI4CSM project that has received funding
	within the ECSEL JU in collaboration with the European Union’s H2020
	Framework Programme and National Authorities, under grant agreement
	No.101007326. The authors kindly acknowledge the financial support of the COMET K2 –Competence Centers for Excellent Technologies Programme of the Austrian Federal Ministry for Climate Action, Environment, Energy, Mobility, Innovation and Technology (BMK), the Austrian Federal Ministry for Digital and Economic Affairs (BMDW), the Province of Styria (Dept. 12) and the Styrian Business Promotion Agency (SFG), under the authorized management of the Austrian Research Promotion Agency (FFG).}
\thanks{\authorrefmark{1}Virtual Vehicle Research GmbH, Graz, Austria {\tt\small karin.tieber@v2c2.at}}%
\thanks{\authorrefmark{2}Institute of Automation and Control, Graz University of Technology, Graz, Austria}%
}

\newcommand{\state}{\boldsymbol{x}}
\newcommand{\sign}{\text{sign}}
\newcommand{\lwb}{\lambda}

\newcommand{\front}{{}_\text{f}}
\newcommand{\lead}{{}_\text{l}}

\newcommand{\veh}{{}_\text{veh}}
\newcommand{\lwbv}{{\lambda}_\text{veh}}

\usepackage{algorithmicx}
\input{algpseudocode.sty}

\begin{document}

\maketitle
\copyrightnotice
\thispagestyle{empty}
\pagestyle{empty}

\begin{abstract}
	When designing path tracking controllers for car-like vehicles, two main aspects are the tracking performance and the characteristics of the actuation signal. Our work bases on an existing variable structure controller that was designed with the geometrically optimal solution of a Dubins car, but with chattering on the output.
	In this contribution, we extend this approach to achieve an actuation signal that is $n$-smooth and differentially bounded. While the global stability under matched disturbances is maintained, the finite time reaching behavior is exchanged for asymptotic convergence. With the $n$-smooth output, $n$ new parameters are introduced, weighing the control characteristics between the tracking performance and the magnitude of the steering angles derivatives.
	The controller is also evaluated in simulation, demonstrating the tuning capabilities, as well as the reaching and tracking behavior. 
	The main contribution of this work is a control law designed to produce a smooth steering angle with implicit satisfaction of bounds on its derivatives. 
\end{abstract}

\section{INTRODUCTION}
Tracking a path is a basic task for many mobile robots. 
It is characteristically for tracking problems with a car-like vehicle, that even the simplest plant models like the Dubins car represent a nonlinear system that can only be approximated with linear equations in a small domain. This is why many path tracking control solutions are designed taking into account these nonlinearities as summarized in surveys \cite{Dominguez2016,Paden2016,Yao2020}.
In \cite{robopt}, a control law based on the optimal solution of a Dubins car reaching a straight reference path was presented. However, this first order sliding mode control produces chattering on the steering angle of the vehicle that is not acceptable in real car-like vehicles.

It is an important aspect when steering car-like vehicles, that the actuation signal is favorable for the actuation system and the driving comfort.
In \cite{Cumali2019}, a nonlinear control law is designed to provide a smooth steering output, which is necessary for a descent steering-feel. 
Other path tracking controllers are extended to achieve certain properties concerning the actuation system. For example to reduce the rack force \cite{Xiaodong2019} or to provide a descent torque feedback to the driver \cite{Loof2019}. The introduced constraints on the steering signal affect the stability and tracking performance of the controllers.
For example in \cite{Klauer2020}, the actuation dynamics are considered by a nonlinear plant system inversion.
With extensions for disturbance compensation and actuation constraints, multiple parameters arise, that can only be selected experimentally and satisfaction of the constraints is not guaranteed.

There are different approaches to deal with input and state constraints while maintaining guaranteed stability. In \cite{Nguyen2021}, fuzzy Lyapunov functions are used to deal with the constraints. In \cite{Xu2020} the path tracking task is formulated as an LQR problem and extended by control saturation and preview of the curvature of the reference path. In \cite{Hu2020}, constraints on the steering signal are introduced by adding barrier functions.\par
An interesting solution to both the consideration of non-linearities and the generation of a smooth control output, is higher order sliding mode control (HOSM). It offers general statements regarding stability and the region of convergence, while at the same time it is possible to increase the smoothness of the control output (i.e. the steering angle) to arbitrary classes. 
In \cite{conditioned_supertwist} and \cite{Incremona2017}, the design of higher order sliding mode control laws with saturated control and state variables is presented. However, with these approaches, it is difficult to cope with non-linearities in the plant system in such a way that global stability is maintained and the control parameters are appropriately chosen for a large domain.\par


The contribution of this paper is the extension of the controller based on Dubins optimal solution, presented in \cite{robopt}, for usage in car-like vehicles.
With the extension, the steering angle is smooth and $n$ derivatives of the steering angle exist and are continuous (the steering angle is \mbox{$C^n$-smooth}). Moreover, the magnitude of the steering angle and its derivatives are bounded.
 The controller is globally asymptotically stable and reaches the reference path in close-to shortest distance. The controller has \mbox{$n+2$} parameters with each parameter tuning the trade-off between each two specific objectives such as actuation effort, noise suppression and robustness to disturbances.\par
The extended controller is designed by first adding state variables to increase the system order and then applying the control law designed in \cite{robopt} to the new variables.
These new state variables are defined as the states of a fictive trailer system. As a result, a fictive lead wheel is steered to the reference path while satisfying the constraint of limited curvature. From kinematic considerations we can show that the resulting steering angle of the real vehicle as well as its derivatives are bounded. While it is easy to see that the rear wheel of the fictive trailer system will asymptotically converge to a straight reference path, we also show how this convergence is achieved in a curved reference path (i.e. the rear wheel of the vehicle is not cutting corners).\par
The paper is structured as follows:
In Sec.~\ref{sec:dubin_ext}, the original control problem is described and then extended step-by-step to finally obtain a $C^n$-smooth steering angle.
Sec.~\ref{sec:paramterization} describes how to choose the control parameters to meet different constraints and objectives. In Sec.~\ref{sec:evaluation}, the controller is evaluated in simulation with a kinetic single track model and actuation delay. The controller is also compared to other control solutions. In Sec.~\ref{sec:sota_placment}, the control design methodology is approached from different perspectives of control design to clarify the classification of the presented control solution.
Sec.~\ref{sec:conclusion} summarizes and concludes the work.

\section{Application of the Dubins car control on car-like vehicles}\label{sec:dubin_ext}
\subsection{Problem}
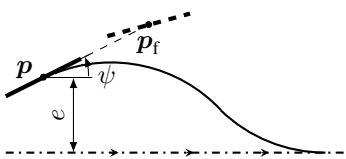
\begin{figure}[b]
	\centering
	\begin{tikzpicture}[]

\def\lengthlabel[#1](#2)(#3)(#4:#5)(#6)
    {\draw (#2)--++(#4+90:#5+0.1);
     \draw (#3)--++(#4+90:#5+0.1);
    \draw (#2)--++(#4+90:#5-0.1);
     \draw (#3)--++(#4+90:#5-0.1);
    \draw (#2)--++(#4+90:#5)coordinate(l11);
     \draw (#3)--++(#4+90:#5)coordinate(l12);
    \draw [latex-latex](l11)--(l12)node[anchor=south, pos=0.5,rotate=#4]{#6};
 }

\def\centerarc[#1](#2)(#3:#4:#5)
    { \draw[#1] ($(#2)+({#5*cos(#3)},{#5*sin(#3)})$) arc (#3:#4:#5); }

\draw[thick,dash dot] (-2,0) --(2.5,0);
\draw[-stealth] (0.4,0) --(0.5,0);
\draw[-stealth] (-0.6,0) --(-0.5,0);
\draw[-stealth] (1.4,0) --(1.5,0);


\draw (-1.5,1) coordinate (v1) {};
\draw[ultra thick]($(v1)+(-0.5,-0.25)$) -- ($(v1)+(0.5,0.25)$) node[anchor=east, pos=0.5,yshift=1mm]{$\boldsymbol{p}$};
\draw ($(v1)-(0,0.02)$) node{\Huge$\cdot$};
\draw[](v1) -- ($(v1)+(0.6,0.3)$);
\draw (v1) -- ($(v1)+(0.65,0)$);
\draw[latex-] ($(v1)+(0.55,0.28)$) arc(22:0:0.7) node [anchor=west, pos=1] {$\psi$};
\draw [latex-latex] (-1.1,0)--(-1.1,1) node[anchor=south, pos=0.5,rotate=90] {$e$};

\draw[dashed] (v1) --++(1.4,0.7) coordinate(v2);
\draw[ultra thick,dashed]($(v2)+(-0.55,-0.15)$) -- ($(v2)+(0.55,0.15)$) node[anchor=north, pos=0.5]{$\boldsymbol{p}_\text{f}$};
\draw ($(v2)-(0,0.02)$) node{\Huge$\cdot$};


\centerarc[thick](-0.6,-0.8)(43:90+22:2)

\centerarc[thick](2.25,2)(-90:-134:2)

\end{tikzpicture}
	\caption{The rear wheel of a car-like vehicle is represented as a Dubins car, which has to approaching a straight, directed reference path.}
	\label{fig:problem}
\end{figure}

In \cite{robopt}, a path tracking controller for a Dubins car has been designed to approach and track a straight, directed reference path. The control problem is illustrated in Fig. \ref{fig:problem}.
With the state vector $\boldsymbol{x} = \begin{bmatrix}
	e & \psi\end{bmatrix}^T$, the dynamic system is:
\begin{equation}\label{eqn:dynsys}
	\frac{d}{ds}\state = \state' = \begin{bmatrix}\sin \psi \\ \kappa \end{bmatrix}
\end{equation}
Where the curvature $\kappa$ of the Dubins car is the control input and $s$ is the distance traveled by the cars rear wheel. \par 
For a Dubins car, where $\kappa$ is constrained by $|\kappa| < \bar\kappa$, the optimal path for reaching the reference path consists of sections of circles and straight lines. Due to that, the optimal path can be expressed as a first order sliding mode control (as described in \cite{robopt}):
\begin{equation}\label{eqn:cntrllaw} \kappa = \bar\kappa\cdot\sign(\sigma(\boldsymbol{x}))
\end{equation}
Assuming $\cos\psi>0$ (this assumption can also be omitted by a modification of the sliding surface), the switching function (the sliding variable) for an optimal path reads as \cite{robopt}:
\begin{equation*}
	\sigma_\text{opt} = -e - \frac{1-\cos \psi}{\bar\kappa}\cdot \sign(\sin \psi)
\end{equation*}
In \cite{robopt}, this function can be modified to account for matched disturbances in the system (i.e. disturbances acting on $\psi'$). The switching function, extended by the tuning parameter $k_\text{rob}$, then is:

\begin{equation}\label{eqn:sigma}
	\sigma = -e - \frac{1-\cos \psi}{ (1-k_\text{rob})\cdot\bar\kappa}\cdot \sign(\sin \psi)
\end{equation}
With $k_\text{rob}>0$, the controller \eqref{eqn:cntrllaw} becomes robust to matched disturbances.

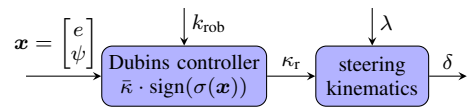
\begin{figure}[h]
	\centering
	\footnotesize
\begin{tikzpicture}

\def\blockwidth{1.5cm}
\def\blockheight{0.8cm}

\node[draw,
rounded corners,
	minimum width = \blockwidth,
    minimum height=\blockheight,
    fill=blue!30,
align=center
] (cntrl) at (0,0){Dubins controller\\$\bar\kappa\cdot\text{sign}(\sigma(\boldsymbol{x}))$};

\node[draw,
rounded corners,
	minimum width = \blockwidth,
    minimum height=\blockheight,
    fill=blue!30,
align=center
] (subst) at ($(cntrl)+(2.5,0)$){steering\\kinematics};

\draw[-stealth] ($(cntrl.west)+(-1,0)$)--(cntrl.west)
      node[pos=0.4,above]{$\boldsymbol{x}=\begin{bmatrix}e\\ \psi\end{bmatrix}$};
\draw[-stealth] (cntrl.east)--(subst.west)
      node[midway,above]{$\kappa_\text{r}$};
\draw[-stealth] (subst.east)--($(subst.east)+(0.5,0)$)
      node[midway,above]{$\delta$};

\draw[stealth-] (cntrl.north)--++(0,0.5) node[pos=0.6,anchor=west]{$k_\text{rob}$};

\draw[stealth-] (subst.north)--++(0,0.5) node[pos=0.6,anchor=west]{$\lambda$};

\end{tikzpicture}
	\caption{Applying the Dubins controller on a car-like vehicle}
\label{fig:cntrl_chart01}
\end{figure}

In a car-like vehicle, the control input is not $\kappa$, but the steering angle $\delta$. For a kinematic single track model, there is a direct relation between steering angle and curvature. The curvature of the path traveled by the rear wheel is:
\begin{equation}\label{eqn:kappar}
	\kappa_\text{r} = \psi' = \frac{\tan(\delta)}{\lwb}
\end{equation}
Where the subscript '$\text{r}$' denotes the rear wheel (the front wheel will be denoted with subscript '$\text{f}$' later) and $\lwb$ is the wheelbase.
Consequently, we can apply the control law \eqref{eqn:cntrllaw} on a car-like vehicle as illustrated in Fig.~\ref{fig:cntrl_chart01}.
The resulting steering angle then is:
\begin{equation}
	\delta = \arctan\left(\bar\kappa\cdot \lwb\cdot\sign(\sigma(\boldsymbol{x}))\right)
\end{equation}
In real car-like vehicles, the steering angle must be continuous and the steering rate is constrained by the actuating system. Therefore, in the next subsection, we will extend the system in order to obtain a control law where the steering rate is constrained and then further extend the system to constrain the 2${}^\text{nd}$ derivative and subsequently the $n^\text{th}$ derivative of the steering angle.

\subsection{Extending the system}\label{sec:ext1}
To achieve a continuous steering angle, we add a $3^\text{rd}$ state to the dynamic system \eqref{eqn:dynsys} as follows:
\begin{equation}\label{eqn:x_ext}
	\boldsymbol{x}_\text{e} = \begin{bmatrix}
		e \\ \psi \\ \delta
	\end{bmatrix},\, \boldsymbol{x}_\text{e}' = \begin{bmatrix}
	\sin\psi\\ \tan\delta / \lwb \\ \delta' 
\end{bmatrix}
\end{equation}

%
%

We now have to find a control law for $\delta'$ that achieves $\lim_{s\to\infty}e = 0$ under the constraints:
\begin{enumerate}[label=C\arabic*,leftmargin=30pt]
	\item $|\delta| \leq \bar\delta$ (vehicle constraint)	\label{con:delta}
	\item $|\delta'|\leq \bar\delta'$ (actuation constraint)\label{con:ddelta}
\end{enumerate}
The boundary $\bar\delta'$ actually results from a constraint on the steering rate $\dot\delta=d\delta/dt$ with $\bar\delta' = \bar{\dot\delta} / v$, where $t$ is the time and $v$ is the velocity of the vehicle.

For finding such a solution, we extend the Dubins car to a kinematic single track model, as illustrated in Fig.~\ref{fig:dubinsext}. The next consideration is based on the following hypothesis:
\begin{hyp}
	\label{hyp:first}
	When the curvature $\kappa\front$ of the front wheel $\boldsymbol{p\front}$ is constrained, then there is an invariant set $\Delta$ for $\delta$ and $\delta'$ is constrained for all $\delta\in\Delta$.
\end{hyp}
The invariant set means \mbox{$\delta(t=t_0) \in \Delta \implies \delta(t>t_0) \in \Delta$}.
To steer the vehicle to the reference path while $\kappa\front$ is constrained, we apply the control law for the Dubins car on the front wheel ($\kappa\front= \bar\kappa\front\cdot\sign(\sigma(\boldsymbol{x}\front))$). 
Naturally, the rear wheel will follow the front wheel asymptotically. This approach is described in more detail in the following.\par

Consider a single track model as illustrated in Fig. \ref{fig:dubinsext}. The control error of the front wheel is:
\begin{subequations}\label{eqn:efront}
\begin{align}\label{eqn:efronta}
	\boldsymbol{x\front} &= \begin{bmatrix}e\front \\ \psi\front \end{bmatrix} = \begin{bmatrix}
		e + \sin(\psi)\cdot \lwb \\ \psi + \delta
	\end{bmatrix} = \boldsymbol{x\front}(\boldsymbol{x_\text{e}}) \\
\label{eqn:efrontb}
 \boldsymbol{x}\front'  &= \begin{bmatrix}
 	\sin\psi\front \\ \kappa\front
 \end{bmatrix}\frac{1}{\cos\delta} =  \begin{bmatrix}
	\sin\psi + \cos\psi\cdot\tan\delta \\ \tan\delta/\lwb+\delta'
\end{bmatrix}
\end{align}
\end{subequations}
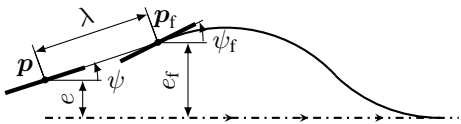
\begin{figure}[b]
\centering
\begin{tikzpicture}[]

\def\lengthlabel[#1](#2)(#3)(#4:#5)(#6)
    {\draw (#2)--++(#4+90:#5+0.1);
     \draw (#3)--++(#4+90:#5+0.1);
    \draw (#2)--++(#4+90:#5-0.1);
     \draw (#3)--++(#4+90:#5-0.1);
    \draw (#2)--++(#4+90:#5)coordinate(l11);
     \draw (#3)--++(#4+90:#5)coordinate(l12);
    \draw [latex-latex](l11)--(l12)node[anchor=south, pos=0.5,rotate=#4]{#6};
 }

\def\centerarc[#1](#2)(#3:#4:#5)
    { \draw[#1] ($(#2)+({#5*cos(#3)},{#5*sin(#3)})$) arc (#3:#4:#5); }

\draw[thick,dash dot] (-3,0) --(2.5,0);
\draw[-stealth] (0.4,0) --(0.5,0);
\draw[-stealth] (-0.6,0) --(-0.5,0);
\draw[-stealth] (1.4,0) --(1.5,0);


\draw (-1.5,1) coordinate (v1) {};
\draw[ultra thick]($(v1)+(-0.5,-0.25)$) -- ($(v1)+(0.5,0.25)$) node[anchor=south, pos=0.6]{$\boldsymbol{p}_\text{f}$};
\draw ($(v1)-(0,0.02)$) node{\Huge$\cdot$};
\draw[](v1) -- ($(v1)+(0.6,0.3)$);
\draw (v1) -- ($(v1)+(0.65,0)$);
\draw[latex-] ($(v1)+(0.55,0.28)$) arc(22:0:0.7) node [anchor=west, pos=1] {$\psi_\text{f}$};
\draw [latex-latex] (-1.1,0)--(-1.1,1) node[anchor=south, pos=0.5,rotate=90] {$e_\text{f}$};

\centerarc[thick](-0.6,-0.8)(43:90+22:2)
\centerarc[thick](2.25,2)(-90:-134:2)

\draw(v1)--(-3,0.5) coordinate(v2);
\draw[ultra thick]($(v2)+(-0.53,-0.17)$) -- ($(v2)+(0.53,0.17)$) node[anchor=south, pos=0.25]{$\boldsymbol{p}$};
\draw ($(v2)-(0,0.02)$) node{\Huge$\cdot$};
\draw (v2) -- ($(v2)+(0.75,0)$);
\draw[latex-] ($(v2)+(0.65,0.25)$) arc(22:0:0.7) node [anchor=west, pos=1] {$\psi$};
\draw [latex-latex] (-2.5,0)--(-2.5,0.5) node[anchor=south, pos=0.5,rotate=90] {$e$};

\lengthlabel[](v1)(v2)(20:0.4)($\lambda$);

\end{tikzpicture}
\caption{The Dubins car approaching a straight, directed reference path.}
\label{fig:dubinsext}
\end{figure}


Where $s\front$ is the path traveled by the front wheel and $ds/ds\front = \cos\delta$.
We assume $\cos\delta > 0$.\par
%
%
Based on \eqref{eqn:efront} we can now observe three properties of the single-track model and proof hypothesis \ref{hyp:first}.\\
\begin{proof}
\begin{itemize}
	\item When the front wheel is steered to the reference path ($\boldsymbol{x}\front = 0)$ then the rear wheel will asymptotically converge to the reference path. From $e\front = e + \sin\psi\cdot\lwb$ and $e'=\sin\psi$ we get:
	\begin{equation}\label{eqn:pt1}
		e' = (e\front - e) / \lwb
	\end{equation}
	This relation corresponds to a first order lag element.
	\item There is an algebraic relation between $\delta'$ and the curvature of the front wheel $\kappa\front$:
	\begin{equation}\label{eqn:ddelta_kappaf}
		\delta' = \kappa\front/\cos\delta - \tan\delta/\lwb
	\end{equation}
	 Thus, for constrained $\kappa\front$ and $\delta$, $\delta'$ will be constrained. 
	\item When $\kappa\front$ is constrained with $|\kappa\front| \leq \bar\kappa\front$ then there is an invariant set $\Delta$ for $\delta$. From \eqref{eqn:ddelta_kappaf} we get:
	\begin{align*}
		\sin\delta/\lambda > \kappa\front &\iff \delta' < 0\text{ and}  \\
		\sin\delta/\lambda \leq \kappa\front &\iff \delta' \geq 0, 
	\end{align*}
	and therefore,
	\begin{equation}\label{eqn:kappaf_constr}
		\Delta = [-\arcsin{\bar\kappa\front\cdot \lwb}, \arcsin{\bar\kappa\front\cdot \lwb}]
	\end{equation}
is an invariant set of $\delta$.
\end{itemize}\vspace{-8pt}\end{proof}

\begin{figure}[b]
	\centering
	\footnotesize
\begin{tikzpicture}

\def\blockwidth{1.5cm}
\def\blockheight{0.8cm}

\node[draw,
rounded corners,
	minimum width = \blockwidth,
    minimum height=\blockheight,
    fill=blue!30,
align=center
] (cntrl) at (0,-3){Dubins controller\\$\bar\kappa_\text{f}\cdot\text{sign}(\sigma(\boldsymbol{x_\text{f}}(\boldsymbol{x}_\text{e}))$};

\node[draw,
rounded corners,
	minimum width = \blockwidth,
    minimum height=\blockheight,
    fill=blue!30,
align=center
] (subst) at ($(cntrl)+(2.7,0)$){steering\\kinematics};

\draw[-stealth] ($(cntrl.west)+(-0.7,0)$)--(cntrl.west)
      node[pos=0.1,above]{$\boldsymbol{x_\text{e}}=\begin{bmatrix}e\\ \psi\\ \delta\end{bmatrix}$};
\draw[-stealth] ($(cntrl.east)+(0,0.2)$)--($(subst.west)+(0,0.2)$)
      node[midway,above]{$\kappa_\text{f}$};
\draw[-stealth] (subst.east)--($(subst.east)+(0.5,0)$)
      node[midway,above]{$\delta'$};
\draw[-stealth] ($(cntrl.west)+(-0.5,0)$)|-++(2,-0.6)-|($(subst.west)+(-0.5,-0.3)$)--($(subst.west)+(0,-0.3)$)
node[pos=0.5, above]{$\delta$};

\draw[stealth-] (cntrl.north)--++(0,0.5) node[pos=0.6,anchor=west, align=left]{$k_\text{rob}$, $\lambda$};
\draw[stealth-] (subst.north)--++(0,0.5) node[pos=0.6,anchor=west]{$\lambda$};

\end{tikzpicture}
	\caption{Applying the Dubins controller on the front wheel of a car-like vehicle.}
\label{fig:cntrl_chart02}
\end{figure}
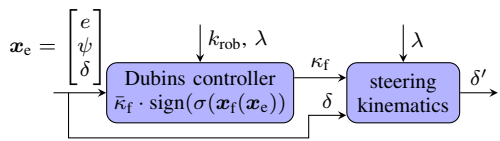

We apply the control law for the Dubins car as illustrated in Fig.~\ref{fig:cntrl_chart02}.
This control will steer $\boldsymbol{x\front}$ to $0$ in finite time (as shown in \cite{robopt}) and thus $e$ will asymptotically converge to $0$.
The control law of the extended system is:
\begin{subequations}\label{eqn:ext1}
	\begin{align}
	\delta' &= \kappa\front / \cos\delta - \tan\delta/\lwb\\
	\label{eqn:ext1b}
	\kappa\front &= \bar\kappa\front \cdot \sign(\sigma(\boldsymbol{x}\front))\\
	\begin{split}\label{eqn:sigmaf}
	\sigma(\boldsymbol{x}\front(\boldsymbol{x}_\text{e})) &= -e- \sin\psi\cdot\lwb\\&-\frac{1-\cos(\psi+\delta)}{(1-k_\text{rob})\cdot\bar\kappa\front}\cdot\sign(\psi+\delta)\end{split}
\end{align}
\end{subequations}

This control law has the following properties:
\begin{itemize}
	\item the tracking error of the rear wheel $e$ will converge to the tracking error of the front wheel $e\front$, which will reach $0$ in finite time
	\item the steering angle $\delta$ has an invariant set $\Delta \ni 0$
	\item the derivative of the steering angle $\delta'$ is constrained
\end{itemize}
The boundaries of $\Delta$ and $\delta'$ will be discussed in Sec.~\ref{sec:paramterization}.
Additionally, it was shown in \cite{robopt}:
\begin{itemize}
	\item when $k_\text{rob} = 0$, the reaching path of the front wheel is optimal under the constraint $|\kappa\front| < \bar\kappa\front$
	\item when $k_\text{rob} \geq \kappa_\text{d}/\bar\kappa\front$, the system \eqref{eqn:efrontb} is robust to disturbances $\kappa_\text{d}$ acting on $d\psi\front/ds\front$ (i.e. $d\psi\front/ds\front = \kappa\front + \kappa_\text{d})$. 
\end{itemize}
Note that the constraint $|\kappa\front| < \bar\kappa\front$ has been defined for the control design and does not origin from the problem itself. The deviation from an optimal solution to the actual problem is evaluated in Sec.~\ref{sec:benchmark}.

\subsection{Extending to $C^1$ smooth output}\label{sec:secondsmooth}
In the previous section, we extended the Dubins car (one wheel with constrained curvature) to a single track model (an additional wheel in front of the first wheel). By applying the control law to the front wheel instead of the rear wheel, we achieved a continuous steering angle with a constrained steering rate. However, in a real vehicle, also the steering rate must be continuous and the steering acceleration $\ddot{\delta}$ is constrained by inertia and requirements regarding comfort. Thus we introduce a constraint additionally to constraints \labelcref{con:delta,con:ddelta}:

\begin{enumerate}[label=C\arabic*]
	\setcounter{enumi}{2}	
	\item $|\delta''| < \bar\delta''$ (inertia constraint)\label{con:dddelta}
\end{enumerate}
\par
\begin{figure}[b]
	\centering
	\begin{tikzpicture}[]
 \def\vehicle[#1]#2(#3)#4(#5,#6,#7,#8)(#9)
  {\node [draw, #1, shape=rectangle, minimum width=#8*0.3cm, minimum height=#8*0.1cm,rotate=#6, rounded corners=2pt,fill, fill opacity=0.5] (#9rear) at (#3) {};
\node [draw, #1, shape=rectangle, minimum width=#8*0.3cm, minimum height=#8*0.1cm,rotate=#6+#7, rounded corners=2pt,fill, fill opacity=0.5](#9front)at ($(#9rear)+(#6:#5)$) {};
\draw[#1] (#9front.center)--(#9rear.center)coordinate[pos=0](#9cg){};}

 \def\vehiclelead[#1]#2(#3)#4(#5,#6,#7,#8)(#9)
  {\node [#1, shape=rectangle, minimum width=#8*0.3cm, minimum height=#8*0.1cm,rotate=#6, rounded corners=2pt, fill opacity=0.5] (#9rear) at (#3) {};
\node [draw, #1, shape=rectangle, minimum width=#8*0.3cm, minimum height=#8*0.1cm,rotate=#6+#7, rounded corners=2pt,fill, fill opacity=0.5](#9front)at ($(#9rear)+(#6:#5)$) {};
\draw[#1] (#9front.center)--(#9rear.center)coordinate[pos=0](#9cg){};}

 \def\arclabel[#1]#2(#3)#4(#5,#6,#7,#8)(#9)
 {\draw[#1](#3)--++(#5:#7);
\draw[#1](#3)--++(#6:#7);
\draw (#3) ++(#5:#8) arc (#5:#6:#8)node[pos=0.5,xshift=-0.2cm,yshift=-0.1cm]{#9};}

\def\centerarc[#1](#2)(#3:#4:#5)
    { \draw[#1] ($(#2)+({#5*cos(#3)},{#5*sin(#3)})$) arc (#3:#4:#5); }

\def\lengthlabel[#1](#2)(#3)(#4:#5)(#6)
    {\draw (#2)--++(#4+90:#5+0.1);
     \draw (#3)--++(#4+90:#5+0.1);
    \draw (#2)--++(#4+90:#5-0.1);
     \draw (#3)--++(#4+90:#5-0.1);
    \draw (#2)--++(#4+90:#5)coordinate(l11);
     \draw (#3)--++(#4+90:#5)coordinate(l12);
    \draw [latex-latex](l11)--(l12)node[anchor=south, pos=0.5,rotate=#4]{#6};
 }

\def\markdot(#1) 
{
\draw ($(#1)-(0,0.02)$) node{\Huge$\cdot$};
}


\def\psiveh{10}
\def\deltaveh{30}
\vehicle[](0,0)(2,\psiveh,\deltaveh,3)(veh);
\vehiclelead[dotted](vehfront.center)(2.2,\psiveh+\deltaveh,29,3)(lead);

\markdot(vehrear.center)
\draw(vehrear.center)node[anchor=south,yshift=3]{$\boldsymbol{p}$};
\markdot(vehfront.center)
\draw(vehfront.center)node[anchor=west,yshift=-6]{$\boldsymbol{p}_\text{f}$};
\markdot(leadfront.center)
\draw(leadfront.center)node[anchor=west]{$\boldsymbol{p}_\text{l}$};

\arclabel[](vehrear.center)(0,\psiveh,0.8,0.7)();
\draw (vehrear.center) node[anchor=west,yshift=8,xshift=10]{$\psi$};

\arclabel[](vehfront.center)(\psiveh,\psiveh+\deltaveh,0.9,0.8)();
\draw (vehfront.center) node[anchor=west,yshift=8,xshift=10]{$\delta$};
\arclabel[](leadfront.center)(\psiveh+\deltaveh,\psiveh+\deltaveh+26.5,0.9,0.8)();
\draw (leadfront.center) node[anchor=west,yshift=14,xshift=4]{$\delta_\text{l}$};

\lengthlabel[](vehrear.center)(vehfront.center)(\psiveh:-0.8)($\lambda$);
\lengthlabel[](leadrear.center)(leadfront.center)(\psiveh+\deltaveh:0.6)($\lambda_\text{l}$);


\end{tikzpicture}
	\caption{Trailer system with 3 wheels: the rear wheel $p$, front wheel $p\front$ and lead wheel $p\lead$.}
\label{fig:trailersystem}
\end{figure}
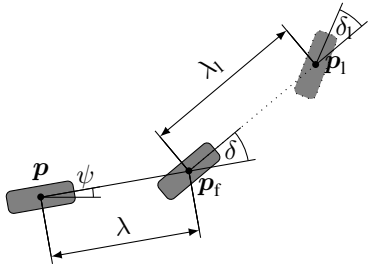

Following the idea from Sec.~\ref{sec:ext1} we extend the system once again by placing an additional wheel in front of the existing wheels. The single track model becomes a trailer system as shown in Fig. \ref{fig:trailersystem}. While the front wheel represents the front axle of the real vehicle, the lead wheel is purely fictive. If we now apply the control law for the Dubins car on the lead wheel (we set the lead wheel curvature to \mbox{$\kappa\lead = \bar\kappa\lead\cdot\sign(\sigma(\boldsymbol{x}\lead))$}), the control error $e$ will be steered to $0$ while the order of continuity is increased (i.e. \labelcref{con:delta,con:ddelta,con:dddelta} will hold). This intuitive result will be derived in the remainder of this section.

\subsubsection{System definition}
The control error of the lead wheel is defined as:
\begin{equation}\label{eqn:xlead}
	\begin{aligned}
				\boldsymbol{x}\lead = \begin{bmatrix}
		e\lead\\\psi\lead
	\end{bmatrix} &= \begin{bmatrix}
		e\front + \sin\psi\front\cdot \lambda\lead\\ \psi\front+\delta\lead
	\end{bmatrix}\\ &= \begin{bmatrix}
		e + \sin\psi\cdot \lwb + \sin(\psi + \delta)\cdot \lambda\lead\\ \psi + \delta+\delta\lead
	\end{bmatrix}
\end{aligned}
\end{equation}

The steering angle $\delta\lead$ of the lead wheel represents the additional state variable of the extended system:
\begin{equation}\label{eqn:ext2}
	\boldsymbol{x}_\text{e,2} = \begin{bmatrix}
		e\\\psi\\\delta\\\delta\lead
	\end{bmatrix}, \boldsymbol{x}_\text{e,2}' = \begin{bmatrix}
		\sin\psi\\\tan\delta/\lwb\\ \frac{\tan\delta\lead}{\lambda\lead}\cdot\frac{1}{\cos\delta} - \frac{\tan\delta}{\lwb}\\ 
		\left(\frac{\kappa\lead}{\cos\delta\lead}-\frac{\delta\lead}{\lambda\lead}\right)/\cos\delta
	\end{bmatrix}
\end{equation}

\subsubsection{Relation between $\delta''$ and $\kappa\lead$}
First, we derive the relation between the curvature $\kappa\lead$ where the control law is applied and the actual control signal, which is the change of steering rate along the traveled path $\delta''$. 

The curvature $\kappa\lead$ is:
\begin{equation}\label{eqn:kappalead}
	\kappa\lead = \frac{d\psi\lead}{ds\lead} =\left(\frac{\tan\delta}{\lwb}+\delta' + \delta\lead'\right)\cdot \cos\delta  \cdot \cos\delta\lead
\end{equation}
From the kinematic relation
$\kappa\front = \tan\delta\lead/\lambda\lead$
and equation~\eqref{eqn:efrontb},
the lead wheel steering angle $\delta\lead$ results in:
\begin{equation}\label{eqn:deltalead}
	\tan\delta\lead = \left(\frac{\tan\delta}{\lwb}+\delta'\right)\cdot\\\lambda\lead\cos\delta
\end{equation}
Its derivative is: 
\begin{equation}\label{eqn:ddeltalead}
	\frac{\delta\lead'}{\cos^2\delta\lead} = \left(\delta'' + \frac{\delta'}{\lwb} - \delta'^2\tan\delta\right)\lambda\lead\cos\delta
\end{equation}

Inserting \eqref{eqn:deltalead} and \eqref{eqn:ddeltalead} into \eqref{eqn:kappalead}, we get the relation between $\kappa\lead$ and $\delta''$:
\begin{equation}\label{eqn:dddelta}
	\begin{aligned}
		\delta'' &= \left(\frac{\kappa\lead}{\cos\delta\cos\delta\lead}-\left(\tan\delta/\lwb+\delta'\right)\right)\\&\cdot\left(\left(\tan\delta/\lwb+\delta'\right)^2\lambda\lead^2\cos^2\delta+1\right)/(\lambda\lead\cos\delta)\\
		&-\delta'/\lwb+\delta'^2\tan\delta
		\end{aligned}
\end{equation}

\subsubsection{Satisfaction of constraints \labelcref{con:delta,con:ddelta,con:dddelta}}
If we can show that the curvature of the front wheel $\kappa\front$ is constrained, then \labelcref{con:delta,con:ddelta} will be satisfied as shown in Sec.~\ref{sec:ext1}.

Combining \eqref{eqn:deltalead} and \eqref{eqn:kappalead} we get:
\begin{equation}
	\delta\lead' = \left(\frac{\kappa\lead}{\cos\delta\lead}  - \frac{\tan\delta\lead}{\lambda\lead}\right) /\cos\delta
\end{equation}
Thus, when $\kappa\lead$ is constrained, there is an invariant set $\Delta\lead$ for $\delta\lead$:
\begin{equation}\label{eqn:kappal_constr}
	\Delta\lead = [-\arcsin{\bar\kappa\lead\cdot \lambda\lead}, \arcsin{\bar\kappa\lead\cdot \lambda\lead}]
\end{equation}

Note that $\Delta\lead$ vanishes for $\bar\kappa\lead\cdot \lambda\lead > 1$. We therefore set the requirement:
\begin{equation}
	\lambda\lead \leq 1/\bar\kappa\lead
\end{equation}
With $\delta\lead \in \Delta\lead$, also $\kappa\front = \tan\delta\lead/\lambda\lead$ is constrained, from which follows the satisfaction of \labelcref{con:delta,con:ddelta}.

From \eqref{eqn:dddelta} it immediately follows that $\delta''$ is also constrained for $\delta\in\Delta, \delta\lead \in\Delta$ and $\cos\bar\delta,\cos\bar\delta\lead >0$. Thus \labelcref{con:dddelta} is satisfied.

\subsubsection{Convergence of $e$}
From \eqref{eqn:efrontb} and \eqref{eqn:xlead} we get: 
\begin{equation}\label{eqn:pt1ext}
	e\lead = e\front + \sin(\psi\front)\cdot\lambda\lead = e\front+ e\front'\cdot\\\lambda\lead
\end{equation}
Thus, $e\front$ will converge to $e\lead$ and (as before) $e\front$ will converge to $e$.

\subsubsection{Resulting control law}
\begin{figure}
	\centering
	\footnotesize
\begin{tikzpicture}

\def\blockwidth{1.5cm}
\def\blockheight{0.8cm}

\node[draw,
rounded corners,
	minimum width = \blockwidth,
    minimum height=\blockheight,
    fill=blue!30,
align=center
] (cntrl) at (0,-6){Dubins controller\\$\bar\kappa_\text{l}\cdot\text{sign}(\sigma(\boldsymbol{x_\text{l}}(\boldsymbol{x}_{\text{e},2}))$};

\node[draw,
rounded corners,
	minimum width = \blockwidth,
    minimum height=\blockheight,
    fill=blue!30,
align=center
] (subst) at ($(cntrl)+(3.5,0)$){steering\\ kinematics}; 

\draw[-stealth] ($(cntrl.west)+(-0.4,0)$)--(cntrl.west);
\draw[-stealth] ($(cntrl.east)+(0,0.25)$)--($(subst.west)+(0,0.25)$)
      node[midway,above]{$\kappa_\text{l}$};
\draw[-stealth] (subst.east)--($(subst.east)+(0.5,0)$)
      node[midway,above]{$\delta''$};
\draw[-stealth] ($(cntrl.west)+(-0.3,0)$)|-++(2,-0.6)node[pos=0.1,left]{$\boldsymbol{x_{\text{e},2}}=\begin{bmatrix}e \\ \psi \\ \delta \\ \delta_\text{l}\end{bmatrix}$}
-|($(subst.west)+(-0.7,-0.3)$)--($(subst.west)+(0,-0.3)$)
node[pos=0.5, above]{$\delta$, $\delta_\text{l}$};

\draw[stealth-] (cntrl.north)--++(0,0.5) node[pos=0.6,anchor=west, align=left]{$k_\text{rob}$, $\lambda$, $\lambda_\text{l}$};
\draw[stealth-] (subst.north)--++(0,0.5) node[pos=0.6,anchor=west]{$\lambda$, $\lambda_\text{l}$};

\end{tikzpicture}
	\caption{Applying the Dubins controller on the lead wheel of a fictive trailer system.}
\label{fig:cntrl_chart03}
\end{figure}
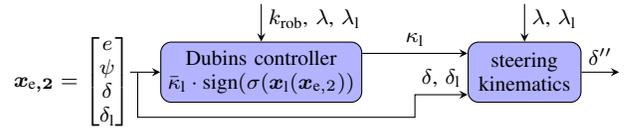

The control law for the Dubins car is now applied as shown in Fig. \ref{fig:cntrl_chart03}. 
The resulting control law is:
\begin{subequations}\label{eqn:control_law}
	\begin{align}
		\begin{split}
		\sigma(\boldsymbol{x}\lead(\boldsymbol{x}_{\text{e},2})) &= -(e+\sin\psi\cdot\lwb + \sin(\psi+\delta)\cdot \lambda\lead)\\&-\frac{1-\cos(\psi+\delta+\delta\lead)}{(1-k_\text{rob})\cdot\bar\kappa\lead}\sign(\sin(\psi+\delta+\delta\lead))\end{split}\\
		\kappa\lead &= \bar\kappa\lead\cdot\sign(\sigma(\boldsymbol{x}\lead))\\
		\delta\lead &= \arctan\left(\left(\frac{\tan\delta}{\lwb}+\delta'\right)\cdot \lambda\lead\cdot\cos\delta\right)\\
		\delta\lead' &= \frac{\kappa\lead}{\cos\delta\cdot\cos\delta\lead}-\frac{\tan\delta}{\lwb}-\delta' \\
		\delta'' &= \frac{\delta\lead'}{\cos^2\delta\lead\cdot\cos\delta\cdot \lambda\lead} - \delta'/\lwb+\delta'^2 \tan\delta
	\end{align}
\end{subequations}
Note that $\delta\lead$ and $\delta'\lead$ are fictive values that relate to each other with $\delta\lead = \int\delta\lead' ds$ only in the ideal, undisturbed case.
The lead wheel state $\boldsymbol{x}\lead(\boldsymbol{x}_{\text{e},2})$ is defined in \eqref{eqn:xlead}. When the reference path is curved, the tracking errors $e\front$ and $e\lead$ differ from the shortest distance between the front and lead wheel and the reference path. When computing $\boldsymbol{x}\lead(\boldsymbol{x}_{\text{e},2})$ from \eqref{eqn:xlead}, the curvature of the reference path will have an influence on the bounds of $\Delta$ and $\Delta\lead$. To prevent this, $e\lead$ is obtained by measuring the shortest distance to the reference path and for $\psi\lead$ the orientation of the path at the point closest to the lead wheel is taken into account. It can be shown that the system then remains globally asymptotically stable for bounded curvature of the reference path.
Note that, to retain the convergence properties, $\lambda$ and $\lambda\lead$ must be greater $0$, as is clear from equation \eqref{eqn:pt1} and \eqref{eqn:pt1ext} respectively. However, for backwards driving ($ds/dt < 0$ i.e. $e' \cdot \dot e < 0$), the dynamics must be reversed to retain convergence, which can be achieved with $\lambda, \lambda\lead < 0$.
\par
The parameters $k_\text{rob}, \lambda$ and $\lambda\lead$ are left to be defined. In Sec.~\ref{sec:paramterization} we will show how to choose them to meet the constraints and target objectives.

\subsection{Generalization to $n$-smooth output}
In the previous sections, we demonstrated how the smoothness of the steering angle $\delta$ can be increased by extending the system order. The approach was the same for $C^{0}$ smoothness and $C^{1}$ smoothness. Thus, this approach can be easily generalized to $C^{n-1}$ smoothness. The control error of the $n^\text{th}$ leading wheel is:
\begin{equation}
	\boldsymbol{x}_n = \begin{bmatrix}
		e_n\\ \psi_n
	\end{bmatrix} = \begin{bmatrix}
	e_{n-1} + \sin\psi_{n-1}\cdot \lambda_n\\
	\psi_{n-1}+\delta_n
\end{bmatrix}
\end{equation}

And the resulting control law is:
\begin{subequations}
	\begin{align}
		\sigma(\boldsymbol{x}_n) &= -e_n - \frac{1-\cos \psi_n}{(1-k_\text{rob})\cdot\bar\kappa_n}\cdot \sign(\sin \psi_n)\\
		\kappa_n &= \bar\kappa_n \cdot \sign(\sigma(\boldsymbol{x}_n))\label{eqn:kappan}
	\end{align}
\end{subequations}

\section{Parametrization}\label{sec:paramterization}
In the presented controller for the $C^1$ smooth steering angle, there appear the parameters $\bar\kappa\lead$, $k_\text{rob}$, $\lambda\lead$ and $\lwb$. Note that the wheelbase of the vehicle $\lambda_\text{veh}$ need not equal $\lwb$ which is the wheelbase of the fictive trailer system. The parameters will be chosen considering mechanical constraints, robustness, noise suppression and steering dynamics (differential bounds of the steering angle).

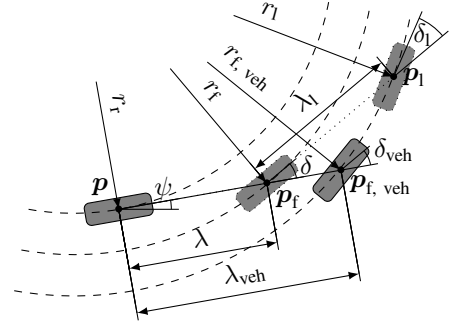
\begin{figure}[h]
	\centering
	\usetikzlibrary{decorations.pathreplacing,calc}
\tikzset{draw half paths/.style 2 args={%
  decoration={show path construction,
    lineto code={
      \draw [#1] (\tikzinputsegmentfirst) -- 
         ($(\tikzinputsegmentfirst)!0.5!(\tikzinputsegmentlast)$);
      \draw [#2] ($(\tikzinputsegmentfirst)!0.5!(\tikzinputsegmentlast)$)
        -- (\tikzinputsegmentlast);
    }
  }, decorate
}}

\begin{tikzpicture}[]
 \def\vehicle[#1]#2(#3)#4(#5,#6,#7,#8)(#9)
  {\node [draw, #1, shape=rectangle, minimum width=#8*0.3cm, minimum height=#8*0.1cm,rotate=#6, rounded corners=2pt,fill, fill opacity=0.5] (#9rear) at (#3) {};
\node [draw, #1, shape=rectangle, minimum width=#8*0.3cm, minimum height=#8*0.1cm,rotate=#6+#7, rounded corners=2pt,fill, fill opacity=0.5](#9front)at ($(#9rear)+(#6:#5)$) {};
\draw[#1] (#9front.center)--(#9rear.center)coordinate[pos=0](#9cg){};}

 \def\vehiclelead[#1]#2(#3)#4(#5,#6,#7,#8)(#9)
  {\node [#1, shape=rectangle, minimum width=#8*0.3cm, minimum height=#8*0.1cm,rotate=#6, rounded corners=2pt, fill opacity=0.5] (#9rear) at (#3) {};
\node [draw, #1, shape=rectangle, minimum width=#8*0.3cm, minimum height=#8*0.1cm,rotate=#6+#7, rounded corners=2pt,fill, fill opacity=0.5](#9front)at ($(#9rear)+(#6:#5)$) {};
\draw[#1] (#9front.center)--(#9rear.center)coordinate[pos=0](#9cg){};}

 \def\arclabel[#1]#2(#3)#4(#5,#6,#7,#8)(#9)
 {\draw[#1](#3)--++(#5:#7);
\draw[#1](#3)--++(#6:#7);
\draw (#3) ++(#5:#8) arc (#5:#6:#8)node[pos=0.5,xshift=-0.2cm,yshift=-0.1cm]{#9};}

\def\centerarc[#1](#2)(#3:#4:#5)
    { \draw[#1] ($(#2)+({#5*cos(#3)},{#5*sin(#3)})$) arc (#3:#4:#5); }

\def\lengthlabel[#1](#2)(#3)(#4:#5)(#6)
    {\draw (#2)--++(#4+90:#5+0.1);
     \draw (#3)--++(#4+90:#5+0.1);
    \draw (#2)--++(#4+90:#5-0.1);
     \draw (#3)--++(#4+90:#5-0.1);
    \draw (#2)--++(#4+90:#5)coordinate(l11);
     \draw (#3)--++(#4+90:#5)coordinate(l12);
    \draw [latex-latex](l11)--(l12)node[anchor=south, pos=0.5,rotate=#4]{#6};
 }

\def\markdot(#1) 
{
\draw ($(#1)-(0,0.02)$) node{\Huge$\cdot$};
}


\def\psiveh{10}
\def\deltaveh{30}
\vehicle[](0,0)(3,\psiveh,41,3)(vehreal);
\vehiclelead[dotted](0,0)(2,\psiveh,\deltaveh,3)(veh);
\vehiclelead[dotted](vehfront.center)(2.2,\psiveh+\deltaveh,29,3)(lead);

\arclabel[](vehrear.center)(0,\psiveh,0.8,0.7)();
\draw (vehrear.center) node[anchor=west,yshift=8,xshift=10]{$\psi$};

\arclabel[](vehfront.center)(\psiveh,\psiveh+\deltaveh,0.5,0.4)();
\draw (vehfront.center) node[anchor=south,yshift=0,xshift=15]{$\delta$};
\arclabel[](vehrealfront.center)(\psiveh,\psiveh+\deltaveh,0.5,0.4)();
\draw (vehrealfront.center) node[anchor=south,yshift=0,xshift=20]{$\delta_\text{veh}$};
\arclabel[](leadfront.center)(\psiveh+\deltaveh,\psiveh+\deltaveh+26.5,0.95,0.85)();
\draw (leadfront.center) node[anchor=west,yshift=16,xshift=4]{$\delta_\text{l}$};

\markdot(vehrear.center)
\draw(vehrear.center)node[anchor=south,yshift=2,xshift=-7]{$\boldsymbol{p}$};
\markdot(vehfront.center)
\draw(vehfront.center)node[anchor=west,yshift=-6]{$\boldsymbol{p}_\text{f}$};
\markdot(leadfront.center)
\draw(leadfront.center)node[anchor=west]{$\boldsymbol{p}_\text{l}$};
\markdot(vehrealfront.center)
\draw(vehrealfront.center)node[anchor=west,yshift=-6]{$\boldsymbol{p}_\text{f, veh}$};

\lengthlabel[](vehrear.center)(vehrealfront.center)(\psiveh:-1.4)($\lambda_\text{veh}$);
\lengthlabel[](vehrear.center)(vehfront.center)(\psiveh:-0.8)($\lambda$);
\lengthlabel[](leadrear.center)(leadfront.center)(\psiveh+\deltaveh:0.25)($\lambda_\text{l}$);

\def\radr{3.45}
\path(vehrear.center)--++(\psiveh+90:\radr)coordinate(cent);
\centerarc[dashed](cent)(-100:-10:\radr);

\def\radf{4}
\centerarc[dashed](cent)(-100:-10:\radf);

\def\radl{4.55}
\centerarc[dashed](cent)(-100:-10:\radl);


\draw[draw half paths={draw=none}{-latex}](cent)--(vehrealrear.center)node[pos=0.6, anchor=south,rotate=-80] {$r_\text{r}$};
\draw[draw half paths={draw=none}{-latex}](cent)--(vehrealfront.center)node[pos=0.6, anchor=south,rotate=-40] {$r_\text{f, veh}$};
\draw[draw half paths={draw=none}{-latex}](cent)--(vehfront.center)node[pos=0.65, anchor=south,rotate=-47] {$r_\text{f}$};
\draw[draw half paths={draw=none}{-latex}](cent)--(leadfront.center)node[pos=0.6, anchor=south,rotate=-15] {$r_\text{l}$};

\end{tikzpicture}
	\caption{Trailer system with 3 wheels. The fictive front wheel is positioned with wheelbase $\lambda$ such that, during stationary cornering, the lead wheel is on the same track as the real front wheel.}
	\label{fig:trailersystem_fic}
\end{figure}

\begin{enumerate}
	\item From the bounds for the steering angle $\bar\delta$, we get the constraint: \begin{equation}\label{eqn:kappalmax}
		\bar\kappa\lead < \frac{\sin\bar\delta}{\lambda^2+\lambda\lead^2\cdot\sin^2\bar\delta}
	\end{equation}
	\item To achieve robustness to disturbances $\kappa_\text{d} \leq \bar\kappa_\text{d}$ acting on the lead wheel curvature $d\psi\lead/ds\lead = \kappa\lead + \kappa_\text{d}$, we get the constraint:
	\begin{equation}
		k_\text{rob}\geq \frac{\kappa_\text{d}}{\bar\kappa\lead}
	\end{equation}
	As demonstrated in \cite{robopt}, the noise suppression decreases with increasing $k_\text{rob}$. To increase the robustness without decreasing the noise suppression, $\bar\kappa\lead$ can be increased.
	\item To eliminate a stationary control error of the front wheel, the fictive wheelbase $\lambda$ is chosen such that the stationary curve radius of the front wheel equals the one of the lead wheel. 
	This is illustrated in Fig. \ref{fig:trailersystem_fic}. This leads to the following constraint:
	\begin{equation}\label{eqn:cornering_constr}
		\lambda^2=\lambda_\text{veh}^2 - \lambda\lead^2
	\end{equation}
	Note that this implies $\lambda\lead < \lwbv$.
	\item The maximum steering rate that can occur is minimal, when the lead wheelbase is $\lambda\lead = 0$. Increasing $\lambda\lead$ will increase the maximum steering rate. This leads to the objective:
	\begin{equation}
		\text{minimize } \lambda\lead
	\end{equation}
	\item The maximum steering acceleration that can occur is minimal when $\lambda\lead = \lambda\lead^* \approx \lambda_\text{veh}$. As $\lambda\lead < \lwbv$, this leads to the objective\footnote{the case $\lambda\lead^* < \lambda\lead < \lambda\veh$ will be treated as exception}:
	\begin{equation}
		\text{maximize } \lambda\lead
	\end{equation}
	Both the maximum steering rate and -acceleration can also be reduced by reducing $\bar\kappa\lead$.
	The relation between the maximum steering rate/-acceleration and $\lambda\lead$ can be computed numerically in form of a static map.
\end{enumerate}

\begin{figure}
	\centering
	\begin{tikzpicture}

\def\lengthlabel[#1](#2)(#3)(#4:#5)(#6)
    {\draw (#2)--++(#4+90:#5+0.1);
     \draw (#3)--++(#4+90:#5+0.1);
    \draw (#2)--++(#4+90:#5-0.1);
     \draw (#3)--++(#4+90:#5-0.1);
    \draw (#2)--++(#4+90:#5)coordinate(l11);
     \draw (#3)--++(#4+90:#5)coordinate(l12);
    \draw [latex-latex](l11)--(l12)node[anchor=south, pos=0.5,rotate=#4]{#6};
 }

\def\width{2.7cm}
\def\height{3cm}
\def\dx{2.5cm}
\def\posy{0cm}
\def\poskappa{0.7}
\def\poskappat{0.4}

\fill[fill=yellow!30] (-\dx,\posy) -- (-\width*\poskappat-\dx,\posy) -- (-\width*\poskappat-\dx,\posy+\height) -- (-\dx,\posy);
\fill[fill=red!30] (-\dx,\posy) -- (-\dx,\height+\posy) -- (-\width*\poskappat-\dx,\height+\posy)--(-\dx,\posy);
\draw (-\width*\poskappat/2-\dx, \posy+\height -0.5cm) node[]{$\bar\delta'$};
\draw (-\width*\poskappat/2-\dx,\posy+0.5cm) node[]{$\bar\delta''$};

\fill[fill=blue!30] (0,\posy) -- (\width,\posy) -- (0,\posy+\height)--(0,\posy)
    node[pos=0.9,right, align=left,yshift=0.2cm]{noise\\supression};
\fill[fill=green!30] (\width,\posy) -- (\width,\posy+\height) -- (0,\posy+\height)--(\width,\posy)
    node[pos=0.1,xshift=0.35cm, yshift=-0.2cm,right,align=right]{matched\\ disturbances};

\fill[fill=yellow!30] (-\dx+0.2cm,\posy+\height)--(-\dx*\poskappa,\posy+\height)--(-\dx*\poskappa,\posy+1cm) -- (-\dx+0.2cm,\posy+1cm);
\fill[fill=red!30] (-\dx+0.2cm,\posy)--(-\dx*\poskappa,\posy)--(-\dx*\poskappa,\posy+1cm) -- (-\dx+0.2cm,\posy+1cm);
\fill[fill=blue!30] (-0.2cm,\posy+\height)--(-\dx*\poskappa,\posy+\height)--(-\dx*\poskappa,\posy+2.5cm) -- (-0.2cm,\posy+2.5cm);
\fill[fill=green!30] (-0.2cm,\posy)--(-\dx*\poskappa,\posy)--(-\dx*\poskappa,\posy+2.5cm) -- (-0.2cm,\posy+2.5cm);

\draw[stealth-,thick] (-\width-\dx/2-0.2cm,0.7cm+\posy)--(-\width-\dx/2-0.2cm,\height-0.7cm+\posy) node [pos=0.5, left]{$\lambda_\text{l}$};
\draw[stealth-,thick] (\width+0.5cm,\posy+0.7cm)--(\width+0.5cm,\posy+\height-0.7cm) node [pos=0.5, right]{$k_\text{rob}$};
\draw[stealth-,thick] (-\dx+0.4cm, \posy+\height+0.5cm) -- (-0.4cm,\posy+\height+0.5cm) node[midway, above]{$\bar\kappa_\text{l}$};

\draw[thick](-\width-\dx/2 +0.1cm,\posy+1cm) -- (-\dx*\poskappa,\posy+1cm);
\draw[thick] (-\dx*\poskappa,\posy+2.5cm) -- (\width+0.2cm,\posy+2.5cm);
\draw[thick] (-\dx*\poskappa,\posy) -- (-\dx*\poskappa,\posy+\height+0.2cm);

\draw[latex-latex, thick] (-\width-\dx/2+0.2cm,0.75cm+\posy)--(-\width-\dx/2+0.2cm,1.25cm+\posy) node[pos=0, left,xshift=0.1cm]{\tiny$+$} node[pos=1, left,xshift=0.1cm]{\tiny$-$};
\draw[-, ultra thick] (-\width-\dx/2+0.2cm,0.9cm+\posy)--(-\width-\dx/2+0.2cm,1.1cm+\posy);

\draw[latex-latex, thick] (\width+0.1cm,\posy+2.25cm) -- (\width+0.1cm,\posy+2.75cm) node[pos=0, right,xshift=-0.1cm]{\tiny$+$} node[pos=1, right,xshift=-0.1cm]{\tiny$-$};
\draw[-, ultra thick] (\width+0.1cm,\posy+2.4cm) -- (\width+0.1cm,\posy+2.6cm);

\draw[latex-latex, thick] (-\dx*\poskappa-0.25cm,\posy+\height+0.1cm) -- (-\dx*\poskappa+0.25cm,\posy+\height+0.1cm) node[pos=0, , anchor=south,yshift=-0.1cm]{\tiny$+$} node[pos=1, anchor=south,yshift=-0.1cm]{\tiny$-$};
\draw[-, ultra thick] (-\dx*\poskappa-0.1cm,\posy+\height+0.1cm) -- (-\dx*\poskappa+0.1cm,\posy+\height+0.1cm);

\filldraw[color=gray] (-2.85,1.0) circle (2pt);
\draw[color=black] (-2.85,1.0) circle (2pt);
\filldraw[color=gray] (0.45,2.5) circle (2pt);
\draw[color=black] (0.45,2.5) circle (2pt);
\filldraw[color=gray] (-1.75,2.5) circle (2pt);
\draw[color=black] (-1.75,2.5) circle (2pt);
\filldraw[color=gray] (-1.75,1) circle (2pt);
\draw[color=black] (-1.75,1) circle (2pt);
\node [rotate=90,fill=white,fill opacity=0.7,inner sep = 0.5] at (-2,0.8) {dynamic};
\node [rotate=90] at (-2,0.8) {dynamic};
\node [rotate=90,fill=white,fill opacity=0.7,inner sep = 0.5] at (-1,0.8) {sensitive};
\node [rotate=90] at (-1,0.8) {sensitive};

\end{tikzpicture}
	\caption{With 3 parameters $\bar\kappa\lead, \lambda\lead, k_\text{rob}$, the trade-off between 4 objectives can be defined. $\lambda\lead$ tunes between high steering rate and high steering acceleration. With $\bar\kappa\lead$, the dynamics in total (both $\bar\delta'$ and $\bar\delta''$) can be decreased at the cost of higher sensitivity to disturbances. $k_\text{rob}$ tunes between higher sensitivity to either matched disturbances or noise suppression.}
\label{fig:trade_off}
\end{figure}
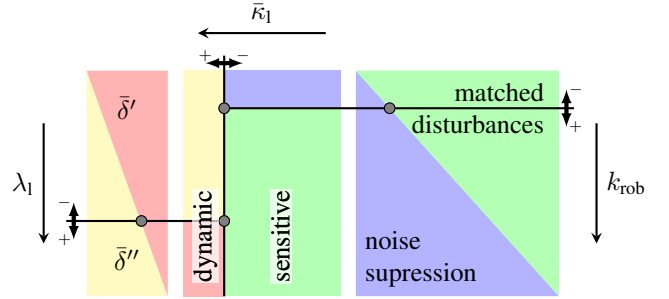

Summarized, the parameter $\lambda\lead$ appears as a parameter to tune the trade-off between maximum steering rate and maximum steering acceleration. Whereas the parameter $\bar\kappa\lead$ can be decreased to decrease both the maximum steering rate and -acceleration. However, the robustness as well as the noise suppression will then decrease. The third parameter $k_\text{rob}$ tunes the trade-off between robustness to unmatched disturbances and noise suppression.
This multi objective trade-off is shown in Fig. \ref{fig:trade_off}.

In the following, we present one possibility to deal with this trade-off.  Additionally to the constraint resulting from the maximum steering angle in equation \eqref{eqn:kappalmax} and the stationary cornering error in equation \eqref{eqn:cornering_constr}, we define the following objectives (sorted by descending priority and assuming constant velocity $\dot v = 0$):
\begin{enumerate}[label=O\arabic*]
	\item the steering rate is constraint (e.g. from the actuation system) $\dot\delta = \delta' \cdot v  \leq \bar{\dot\delta}$\label{o:ddelta}
	\item the steering acceleration is constraint (e.g. from vehicle dynamics\label{o:dddelta} considerations) $\ddot\delta = \delta'' \cdot v^2 \leq \bar{\ddot\delta}$
	\item the maximum curvature $\bar\kappa\lead$ shall be as large as possible to minimize the sensitivity to disturbances\label{o:kappal}
	\item the lead wheel distance $\lambda\lead$ shall be as small as possible to minimize the steering acceleration \label{o:ll}
\end{enumerate}
The parameter $k_\text{rob}$ can be defined afterwards, to tune the controller between robustness to matched disturbances and noise suppression.\par
\begin{figure}
	\centering
	\includegraphics[width=0.5\textwidth, trim=0.1cm 0cm 0 0.5cm,clip]{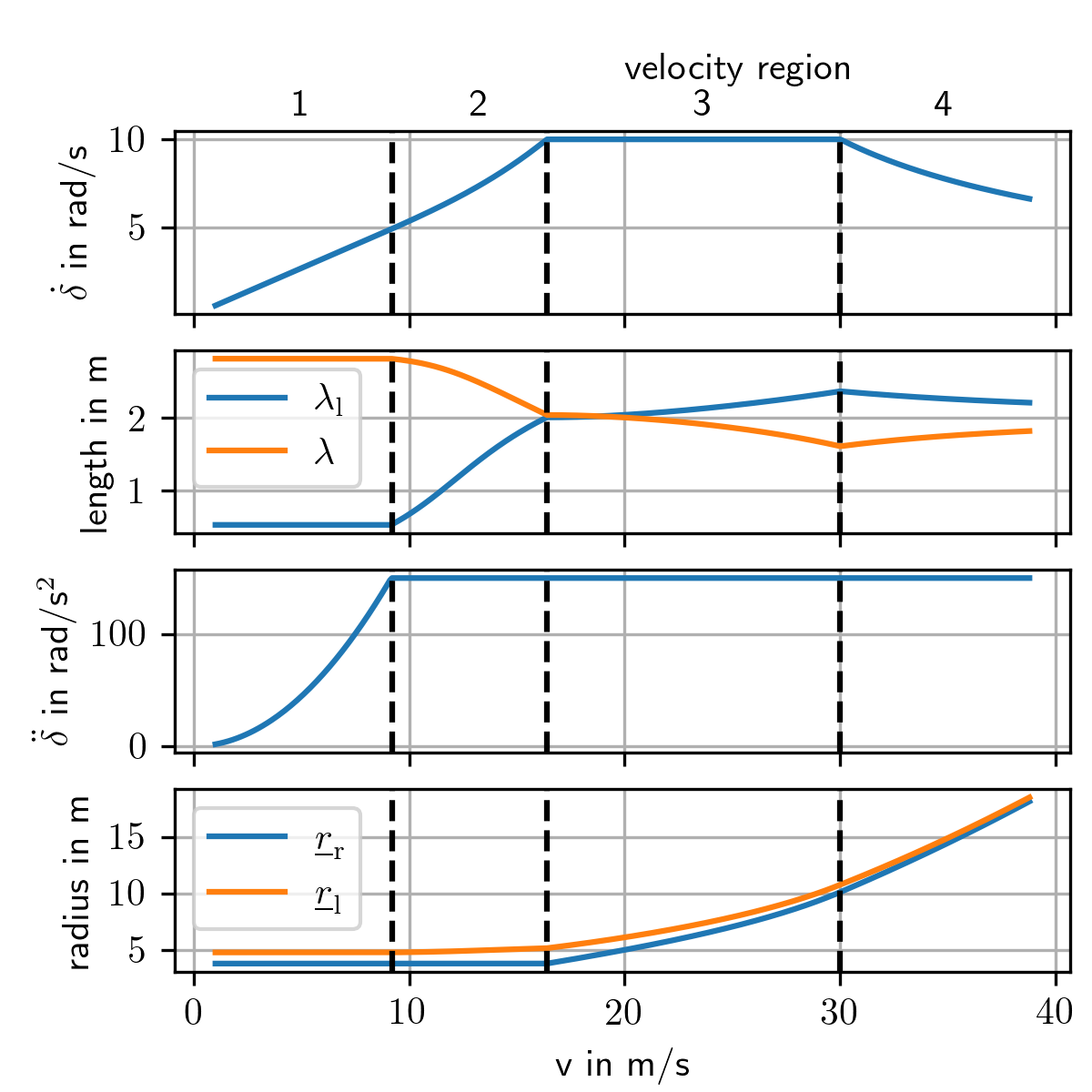}
	\caption{Parameters $\lambda$, $\lambda\lead$ and $\underline r\lead$ and corresponding magnitude of the derivatives of $\delta$ for an optimal, velocity dependent parameter setting.}
\label{fig:param_v}
\end{figure}
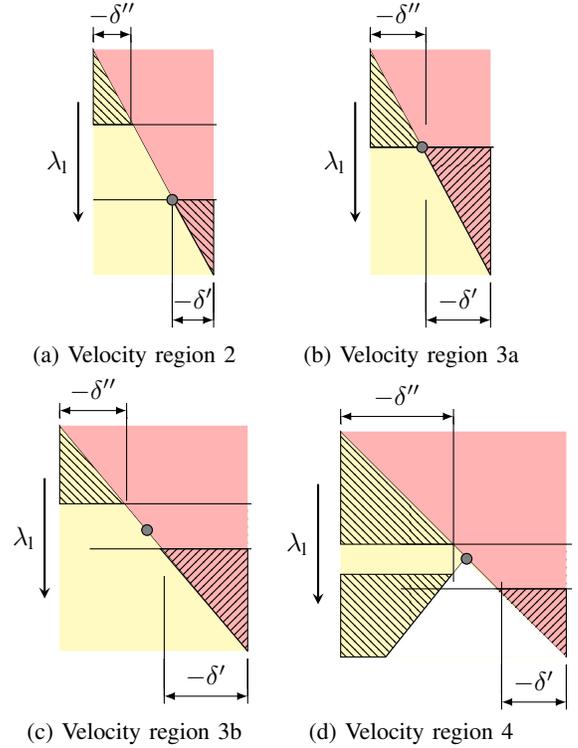
\begin{figure}[h]
	\centering
	\begin{subfigure}[b]{0.2\textwidth}
		\centering
		\begin{tikzpicture}

\def\lengthlabel[#1](#2)(#3)(#4:#5)(#6)
    {\draw (#2)--++(#4+90:#5+0.1);
     \draw (#3)--++(#4+90:#5+0.1);
    \draw (#2)--++(#4+90:#5-0.1);
     \draw (#3)--++(#4+90:#5-0.1);
    \draw (#2)--++(#4+90:#5)coordinate(l11);
     \draw (#3)--++(#4+90:#5)coordinate(l12);
    \draw [latex-latex](l11)--(l12)node[anchor=south, pos=0.5,rotate=#4]{#6};
 }

\def\width{2cm}
\def\height{3cm}
\def\dx{0.1cm}
\def\posy{0cm}
\def\poskappa{0.6}
\def\poskappat{0.8}
\def\llmax{1cm}
\def\llmin{2cm}

\fill[fill=yellow!30,] (-\dx,\posy) -- (-\width*\poskappat-\dx,\posy) -- (-\width*\poskappat-\dx,\posy+\height) -- (-\dx,\posy);
\draw[pattern=north west lines, pattern color=black] (0,0) (-\dx,\posy+\llmin) -- (-\width*\poskappat-\dx,\posy+\llmin) -- (-\width*\poskappat-\dx,\posy+\height) -- (-\dx,\posy);

\fill[fill=red!30] (-\dx,\posy) -- (-\dx,\height+\posy) -- (-\width*\poskappat-\dx,\height+\posy)--(-\dx,\posy);
\draw[pattern=north west lines, pattern color=black] (-\dx,\posy) -- (-\dx,\llmax+\posy) -- (-\llmax*0.8-\dx+0.25cm,\llmax+\posy)--(-\dx,\posy);

\draw[stealth-,thick] (-\width-\dx+0.2cm,0.7cm+\posy)--(-\width-\dx+0.2cm,\height-0.7cm+\posy)node [pos=0.5, left]{$\lambda_\text{l}$};

\draw[](-\width-\dx+0.4cm,\posy+\llmax) -- (-\dx*\poskappa,\posy+\llmax);
\draw[](-\width-\dx+0.4cm,\posy+\llmin) -- (-\dx*\poskappa,\posy+\llmin);
\draw[] (-\dx-\llmax*\poskappat+0.25cm,\posy-0.1cm) coordinate(ddelta_lab) -- (-\dx-\llmax*\poskappat+0.25cm,\posy+1cm);
\draw[] (-\dx-\llmin*\poskappat+0.5cm,\height+\posy+0.1cm) coordinate(dddelta_lab) -- (-\dx-\llmin*\poskappat+0.5cm,\height+\posy-1cm);

\lengthlabel[](-\dx-\width*\poskappat,\height+\posy+0.1cm)(dddelta_lab)(0:0.1)($-\delta''{}$);
\lengthlabel[](-\dx,\posy-0.1cm)(ddelta_lab)(0:-0.5)($-\delta'{}$);

\filldraw[color=gray] (-\llmax*0.8-\dx+0.25cm,\llmax+\posy) circle (2pt);
\draw[color=black] (-\llmax*0.8-\dx+0.25cm,\llmax+\posy) circle (2pt);

\end{tikzpicture}
		\caption{Velocity region 2}
		\label{fig:phase_2}
	\end{subfigure}
	\begin{subfigure}[b]{0.2\textwidth}
	\centering
	\begin{tikzpicture}

\def\lengthlabel[#1](#2)(#3)(#4:#5)(#6)
    {\draw (#2)--++(#4+90:#5+0.1);
     \draw (#3)--++(#4+90:#5+0.1);
    \draw (#2)--++(#4+90:#5-0.1);
     \draw (#3)--++(#4+90:#5-0.1);
    \draw (#2)--++(#4+90:#5)coordinate(l11);
     \draw (#3)--++(#4+90:#5)coordinate(l12);
    \draw [latex-latex](l11)--(l12)node[anchor=south, pos=0.5,rotate=#4]{#6};
 }

\def\width{2cm}
\def\height{3cm}
\def\dx{0.1cm}
\def\posy{0cm}
\def\poskappa{0.6}
\def\poskappat{0.8}
\def\llmax{1.7cm}
\def\llmin{1.7cm}

\fill[fill=yellow!30,] (-\dx,\posy) -- (-\width*\poskappat-\dx,\posy) -- (-\width*\poskappat-\dx,\posy+\height) -- (-\dx,\posy);
\draw[pattern=north west lines, pattern color=black] (0,0) (-\dx,\posy+\llmin) -- (-\width*\poskappat-\dx,\posy+\llmin) -- (-\width*\poskappat-\dx,\posy+\height) -- (-\dx,\posy);

\fill[fill=red!30] (-\dx,\posy) -- (-\dx,\height+\posy) -- (-\width*\poskappat-\dx,\height+\posy)--(-\dx,\posy);
\draw[pattern=north east lines, pattern color=black] (-\dx,\posy) -- (-\dx,\llmax+\posy) -- (-\llmax*0.8-\dx+0.45cm,\llmax+\posy)--(-\dx,\posy);

\draw[stealth-,thick] (-\width-\dx+0.2cm,0.7cm+\posy)--(-\width-\dx+0.2cm,\height-0.7cm+\posy)node [pos=0.5, left]{$\lambda_\text{l}$};

\draw[](-\width-\dx+0.4cm,\posy+\llmax) -- (-\dx*\poskappa,\posy+\llmax);
\draw[](-\width-\dx+0.4cm,\posy+\llmin) -- (-\dx*\poskappa,\posy+\llmin);
\draw[] (-\dx-\llmax*\poskappat+0.5cm,\posy-0.1cm) coordinate(ddelta_lab) -- (-\dx-\llmax*\poskappat+0.5cm,\posy+1cm);
\draw[] (-\dx-\llmin*\poskappat+0.5cm,\height+\posy+0.1cm) coordinate(dddelta_lab) -- (-\dx-\llmin*\poskappat+0.5cm,\height+\posy-1cm);

\lengthlabel[](-\dx-\width*\poskappat,\height+\posy+0.1cm)(dddelta_lab)(0:0.1)($-\delta''{}$);
\lengthlabel[](-\dx,\posy-0.1cm)(ddelta_lab)(0:-0.5)($-\delta'{}$);

\filldraw[color=gray] (-\llmax*0.8-\dx+0.45cm,\llmax+\posy) circle (2pt);
\draw[color=black] (-\llmax*0.8-\dx+0.45cm,\llmax+\posy) circle (2pt);

\end{tikzpicture}
	\caption{Velocity region 3a}
	\label{fig:phase_3a}
	\end{subfigure}
	\begin{subfigure}[b]{0.2\textwidth}
		\centering
		\begin{tikzpicture}

\def\lengthlabel[#1](#2)(#3)(#4:#5)(#6)
    {\draw (#2)--++(#4+90:#5+0.1);
     \draw (#3)--++(#4+90:#5+0.1);
    \draw (#2)--++(#4+90:#5-0.1);
     \draw (#3)--++(#4+90:#5-0.1);
    \draw (#2)--++(#4+90:#5)coordinate(l11);
     \draw (#3)--++(#4+90:#5)coordinate(l12);
    \draw [latex-latex](l11)--(l12)node[anchor=south, pos=0.5,rotate=#4]{#6};
 }

\def\width{2.5cm}
\def\height{3cm}
\def\dx{0.1cm}
\def\posy{0cm}
\def\poskappa{0.5}
\def\poskappat{1.0}
\def\llmax{1.7cm*0.8/\poskappat}
\def\llmin{3cm-1.3cm*0.8/\poskappat}

\fill[fill=yellow!30,] (-\dx,\posy) -- (-\width*\poskappat-\dx,\posy) -- (-\width*\poskappat-\dx,\posy+\height) -- (-\dx,\posy);
\draw[pattern=north west lines, pattern color=black] (0,0) (-\dx,\posy+\llmin) -- (-\width*\poskappat-\dx,\posy+\llmin) -- (-\width*\poskappat-\dx,\posy+\height) -- (-\dx,\posy);

\fill[fill=red!30] (-\dx,\posy) -- (-\dx,\height+\posy) -- (-\width*\poskappat-\dx,\height+\posy)--(-\dx,\posy);
\draw[pattern=north east lines, pattern color=black] (-\dx,\posy) -- (-\dx,\llmax+\posy) -- (-\llmax*\poskappat-\dx+0.2cm,\llmax+\posy)--(-\dx,\posy);

\draw[stealth-,thick] (-\width-\dx-0.2cm,0.7cm+\posy)--(-\width-\dx-0.2cm,\height-0.7cm+\posy)node [pos=0.5, left]{$\lambda_\text{l}$};

\draw[](-\width-\dx+0.4cm,\posy+\llmax) -- (-\dx*\poskappa,\posy+\llmax);
\draw[](-\width-\dx+0.4cm,\posy+\llmin) -- (-\dx*\poskappa,\posy+\llmin);
\draw[] (-\dx-\llmax*\poskappat+0.25cm,\posy-0.1cm) coordinate(ddelta_lab) -- (-\dx-\llmax*\poskappat+0.25cm,\posy+1cm);
\draw[] (-\dx-3cm+1.3cm*0.8*\poskappat+0.35cm,\height+\posy+0.1cm) coordinate(dddelta_lab) -- (-\dx-3cm+1.3cm*0.8*\poskappat+0.35cm,\height+\posy-1cm);

\lengthlabel[](-\dx-\width*\poskappat,\height+\posy+0.1cm)(dddelta_lab)(0:0.1)($-\delta''{}$);
\lengthlabel[](-\dx,\posy-0.1cm)(ddelta_lab)(0:-0.5)($-\delta'{}$);

\filldraw[color=gray] (-\llmax*0.8-\dx-0.25cm,\llmax+\posy+0.25cm) circle (2pt);
\draw[color=black] (-\llmax*0.8-\dx-0.25cm,\llmax+\posy+0.25cm) circle (2pt);

\end{tikzpicture}
		\caption{Velocity region 3b}
	\label{fig:phase_3b}
	\end{subfigure}
	\begin{subfigure}[b]{0.2\textwidth}
	\centering
	\begin{tikzpicture}

\def\lengthlabel[#1](#2)(#3)(#4:#5)(#6)
    {\draw (#2)--++(#4+90:#5+0.1);
     \draw (#3)--++(#4+90:#5+0.1);
    \draw (#2)--++(#4+90:#5-0.1);
     \draw (#3)--++(#4+90:#5-0.1);
    \draw (#2)--++(#4+90:#5)coordinate(l11);
     \draw (#3)--++(#4+90:#5)coordinate(l12);
    \draw [latex-latex](l11)--(l12)node[anchor=south, pos=0.5,rotate=#4]{#6};
 }

\def\width{2cm}
\def\height{3cm}
\def\dx{0.1cm}
\def\posy{0cm}
\def\poskappa{0.25}
\def\poskappat{1.5}
\def\llmax{1.7cm*0.8/\poskappat}
\def\llmin{1.5cm}

\fill[fill=yellow!30,] (-\dx,\posy) -- (-\width*\poskappat-\dx,\posy) -- (-\width*\poskappat-\dx,\posy+\height) -- (-\dx,\posy);
\draw[pattern=north west lines, pattern color=black] (0,0) (-\dx,\posy+\llmin) -- (-\width*\poskappat-\dx,\posy+\llmin) -- (-\width*\poskappat-\dx,\posy+\height) -- (-\dx,\posy);

\fill[fill=red!30] (-\dx,\posy) -- (-\dx,\height+\posy) -- (-\width*\poskappat-\dx,\height+\posy)--(-\dx,\posy);
\draw[pattern=north east lines, pattern color=black] (-\dx,\posy) -- (-\dx,\llmax+\posy) coordinate[](redpat1) -- (-\llmax*\poskappat-\dx,\llmax+\posy)--(-\dx,\posy);

\draw[] (-\dx-1.35cm,\posy+1.3cm) coordinate[](asdf1){} -- (-\dx-\width*1.2,\posy) coordinate[](asdf){};
\fill[fill=white!100] (-\dx,\posy) -- (asdf) -- (asdf1);

\draw[stealth-,thick] (-\width-\dx-1.3cm,0.7cm+\posy)--(-\width-\dx-1.3cm,\height-0.7cm+\posy)node [pos=0.5, left]{$\lambda_\text{l}$};

\draw[](-\width-\dx-0.2cm,\posy+\llmax) -- (-\dx*\poskappa,\posy+\llmax);
\draw[](-\width-\dx-0.2cm,\posy+\llmin) -- (-\dx*\poskappa,\posy+\llmin);
\draw[] (-\dx-\llmax*\poskappat+0.5cm,\posy-0.1cm) coordinate(ddelta_lab) -- (-\dx-\llmax*\poskappat+0.5cm,\posy+1cm);
\draw[] (-\dx-\llmin*\poskappat+0.75cm,\height+\posy+0.1cm) coordinate(dddelta_lab) -- (-\dx-\llmin*\poskappat+0.75cm,\height+\posy-2cm);

\lengthlabel[](-\dx-\width*\poskappat,\height+\posy+0.1cm)(dddelta_lab)(0:0.1)($-\delta''{}$);
\lengthlabel[](-\dx,\posy-0.1cm)(ddelta_lab)(0:-0.5)($-\delta'{}$);


\draw[pattern=north west lines, pattern color=black] (asdf) -- (-\dx-\width*\poskappat, \posy) -- ++(0,1.1cm) -- (-\dx-1.52cm,\posy+1.1cm) -- (asdf);

\filldraw[color=gray] (-\llmax*0.8-\dx-0.6cm,\llmax+\posy+0.4cm) circle (2pt);
\draw[color=black] (-\llmax*0.8-\dx-0.6cm,\llmax+\posy+0.4cm) circle (2pt);

\end{tikzpicture}\vspace{-0.5cm}
	\caption{Velocity region 4}
	\label{fig:phase_4}
	\end{subfigure}
\caption{Parameterizing $\lwb\lead$ to satisfy constraints on $\dot\delta$ and $\ddot\delta$ in different regions of velocity. For increasing velocity, the limits for $\delta'$ and $\delta''$ decrease.}
\label{fig:phases}
\end{figure}

Due to the objectives \labelcref{o:ddelta,o:dddelta}, the optimal parameters vary depending on the velocity. For increasing velocity, the limits for $\delta'$ and $\delta''$ decrease. The resulting optimal parameters $\lambda\lead$ and $\bar\kappa\lead = \frac{1}{\underline{r}\lead}$ are shown in Fig. \ref{fig:param_v}. The effect of decreasing $\max \delta'$ and $\max \delta''$ is different in different regions of velocity. These regions are described in the following and illustrated in Fig. \ref{fig:phases}.

\subsubsection{Velocity region 1}
The velocity is so low that neither $\bar{\dot\delta}$ nor $\bar{\ddot\delta}$ is reached for any parameter setting. Therefore, $\kappa\lead$ is set to the maximum (\labelcref{o:kappal}), which is defined in equation \eqref{eqn:kappalmax}, and $\lwb\lead$ is set to the minimum (\labelcref{o:ll}).

\subsubsection{Velocity region 2}
The constraint for $\ddot\delta$ (\labelcref{o:ddelta}) is reached. To not exceed $\bar{\ddot\delta}$ with increasing velocity, the controller is tuned towards lower $\delta''$ and larger $\delta'$. For this, $\lambda\lead$ is increased.

\subsubsection{Velocity region 3}
Both constraints for $\dot\delta$ (\labelcref{o:ddelta}) and $\ddot\delta$ (\labelcref{o:dddelta}) are reached. To not exceed these constraints with increasing velocity, the controller is tuned towards performance rather than robustness. For this, $\bar\kappa\lead$ is decreased. The distance $\lambda\lead$ is chosen such that both $\dot\delta$ and $\ddot\delta$ are within the constraints. Since $\ddot\delta$ increases faster than $\dot\delta$, $\lambda\lead$ further increases.

\subsubsection{Velocity region 4}
The length $\lambda\lead$ is so large that the minimum of $\delta''$ is transcended. To satisfy \labelcref{o:dddelta} with the maximum possible $\kappa\lead$ (\labelcref{o:kappal}), $\lambda\lead$ is chosen such that $\delta''$ is minimal. \labelcref{o:ddelta} is satisfied anyways.

\section{Evaluation}\label{sec:evaluation} 
The proposed controller is evaluated in three scenarios. First, in Sec.~\ref{subsec:noise}, the controller is simulated in a realistic environment to demonstrate the tracking performance and tuning behavior. Second, in Sec.~\ref{sec:benchmark}, the reaching behavior of the controller is evaluated in a simplified environment to contrast this characteristic to selected other control solutions. Last, in Sec.~\ref{sec:lanechange}, the controller is tested in a lane change scenario to demonstrate the behavior in a usual, realistic urban use case.

\subsection{Noise suppression and robustness}\label{subsec:noise}
\begin{figure}  
	\centering
	\begin{subfigure}[b]{0.5\textwidth}
		\centering
		\includegraphics[width=\textwidth, trim=0cm 0.2cm 0.6cm 1cm,clip]{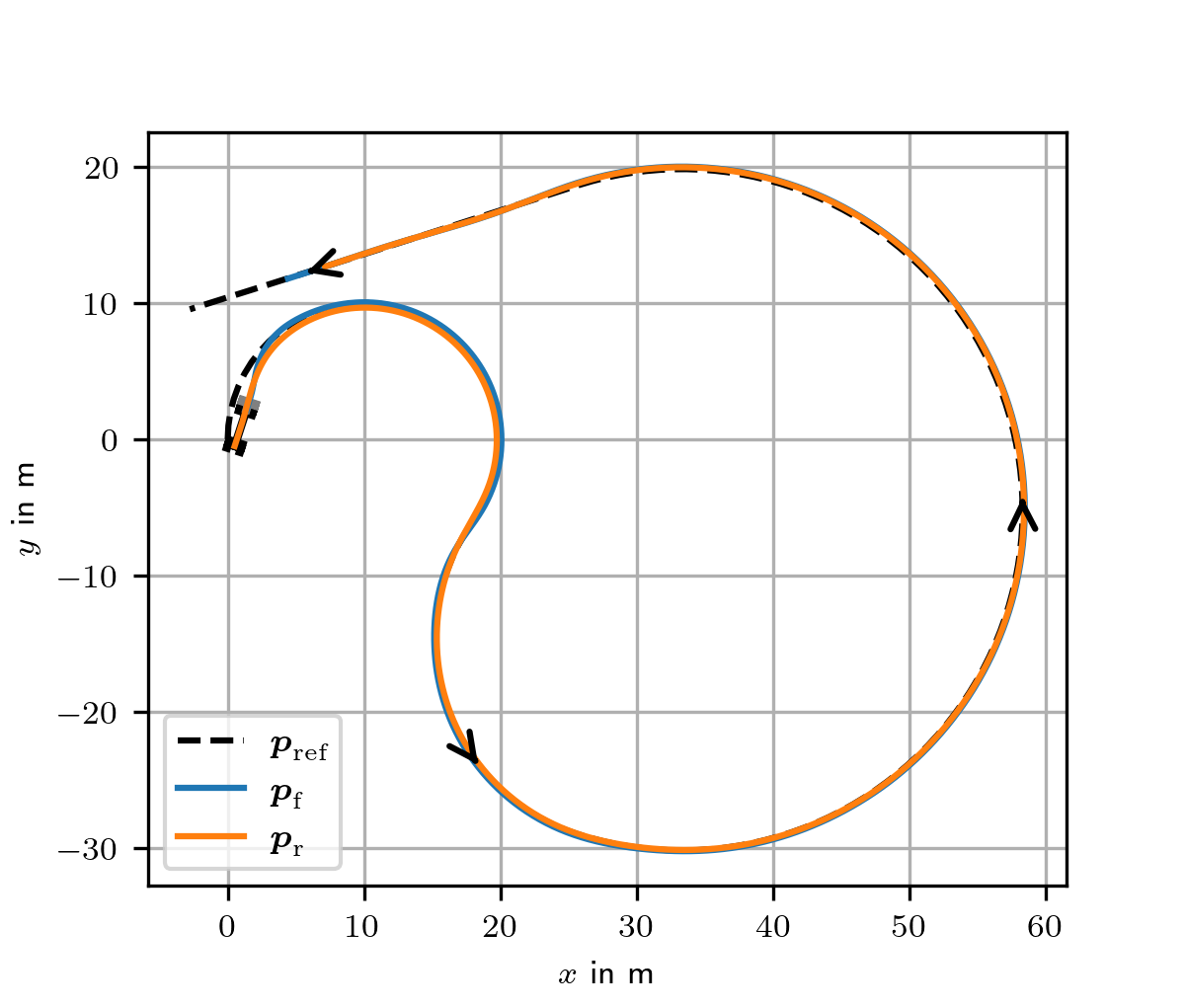}
		\caption{Reference path and path driven by the vehicle}
		\label{fig:simu_p}
	\end{subfigure}
	\begin{subfigure}[b]{0.5\textwidth}
	\centering
	\includegraphics[width=\textwidth, trim=0.3cm 0.2cm 0cm 0cm,clip]{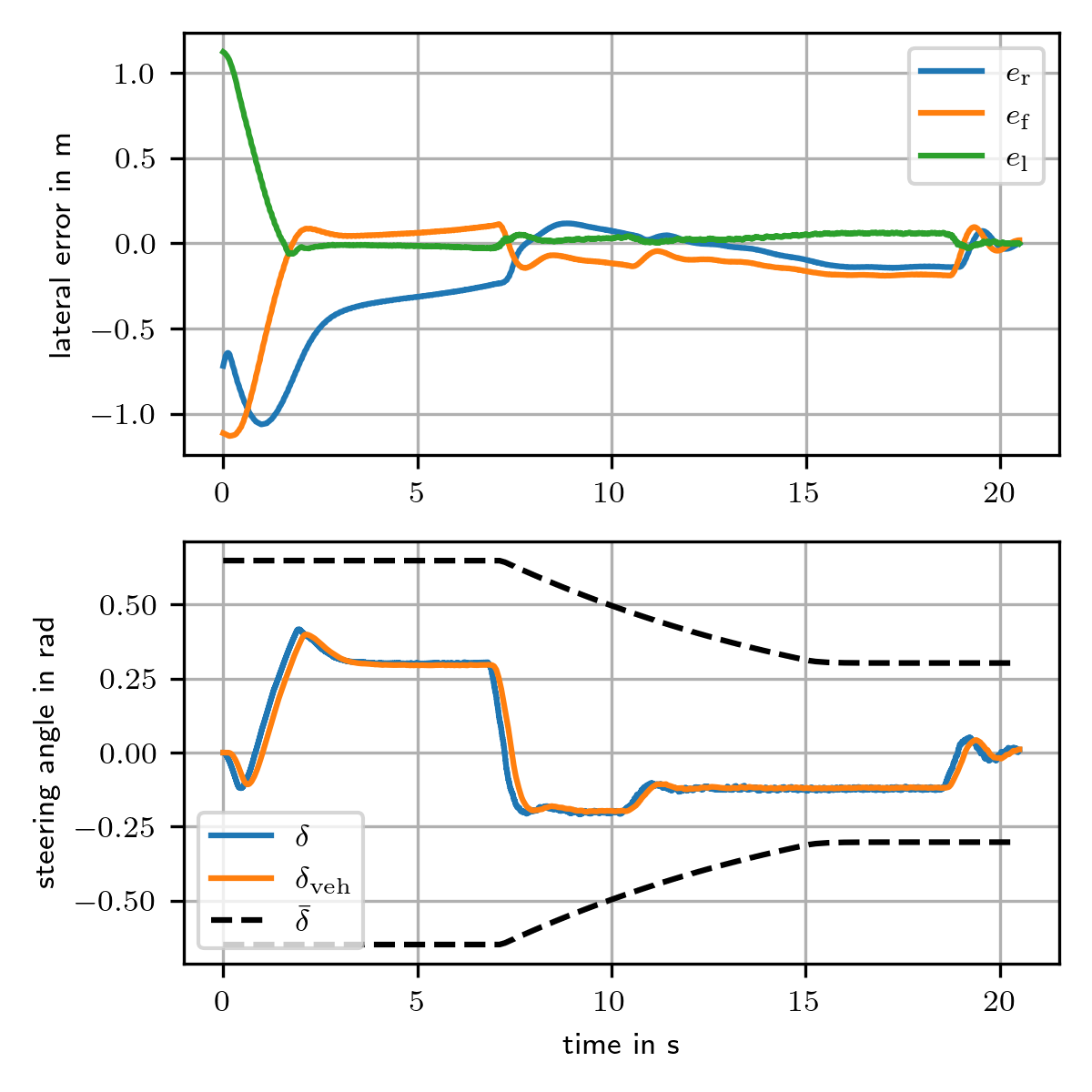}
	\caption{Tracking error and steering angle}
	\label{fig:simu_e}
\end{subfigure}
	\caption{Simulation of the controller with a $C^1$ smooth output under disturbances.}
\label{fig:simu}
\end{figure}

The proposed controller with $C^1$ smooth output is evaluated in simulation. The vehicle is a kinetic single track model with a Pacejka tire model \cite{vehicleDynamics} and a third-order low pass filter for the steering system. The vehicle accelerates from $2.8$ to $12.5$m/s and has to follow a path consisting of circular sections as shown in Fig.~\ref{fig:simu}.

The control parameters are chosen optimal regarding objectives \labelcref{o:ddelta,o:dddelta,o:kappal,o:ll} and $k_\text{rob} = 3$. Additionally, 3 other parameter settings are chosen, where one parameter is varied each:
\begin{itemize}
	\item decrease $\bar\kappa\lead$ by $50\%$
	\item increase $\lambda\lead$ by $300\%$
	\item decrease $k_\text{rob}$ by $33\%$
\end{itemize}

In Fig. \ref{fig:simu}, the control error of the optimal parameter setting is shown. While the control error of the lead wheel $e\lead$ remains below $0.05$m after the reaching phase, the control errors of the front- and rear wheel reach values of up to $0.17$m or $0.31$m respectively. This error consists of two parts: the kinematic tracking error (as the wheels are lagging behind the lead wheel) and the offset induced by the side slip of the vehicle.
The lower diagram of Fig. \ref{fig:simu_e} shows the steering angle $\delta$ that is output by the controller as well as the steering angle $\delta_\text{veh}$ that results from the steering system. The controller is executed at a frequency of $50$Hz. When this frequency is reduced to $20$Hz, the control errors do not change significantly while the steering angle starts to oscillate with an amplitude of $0.012$rad. However, these oscillations are removed by the low pass filter of the steering system.\par

\begin{figure} 
	\centering
	\includegraphics[width=0.5\textwidth, trim=0cm 0cm 0.7cm 0.8cm,clip]{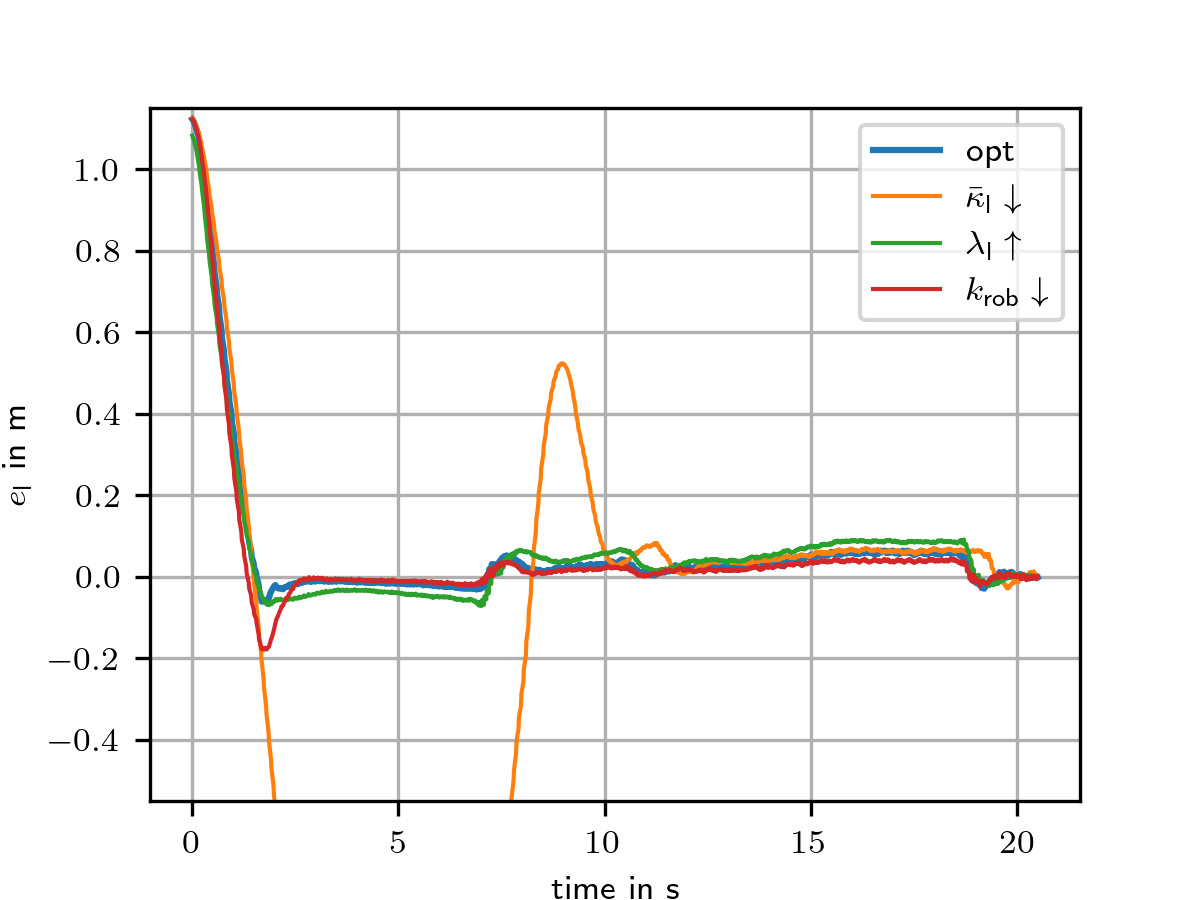}
	\caption{Control error in the scenario shown in Fig. \ref{fig:simu_p} for different parameter settings.}
	\label{fig:el_comp}
\end{figure}

\begin{figure} 
	\centering
	\includegraphics[width=0.4375\textwidth, trim=0.2cm 0.2cm 0.2cm 0.2cm,clip]{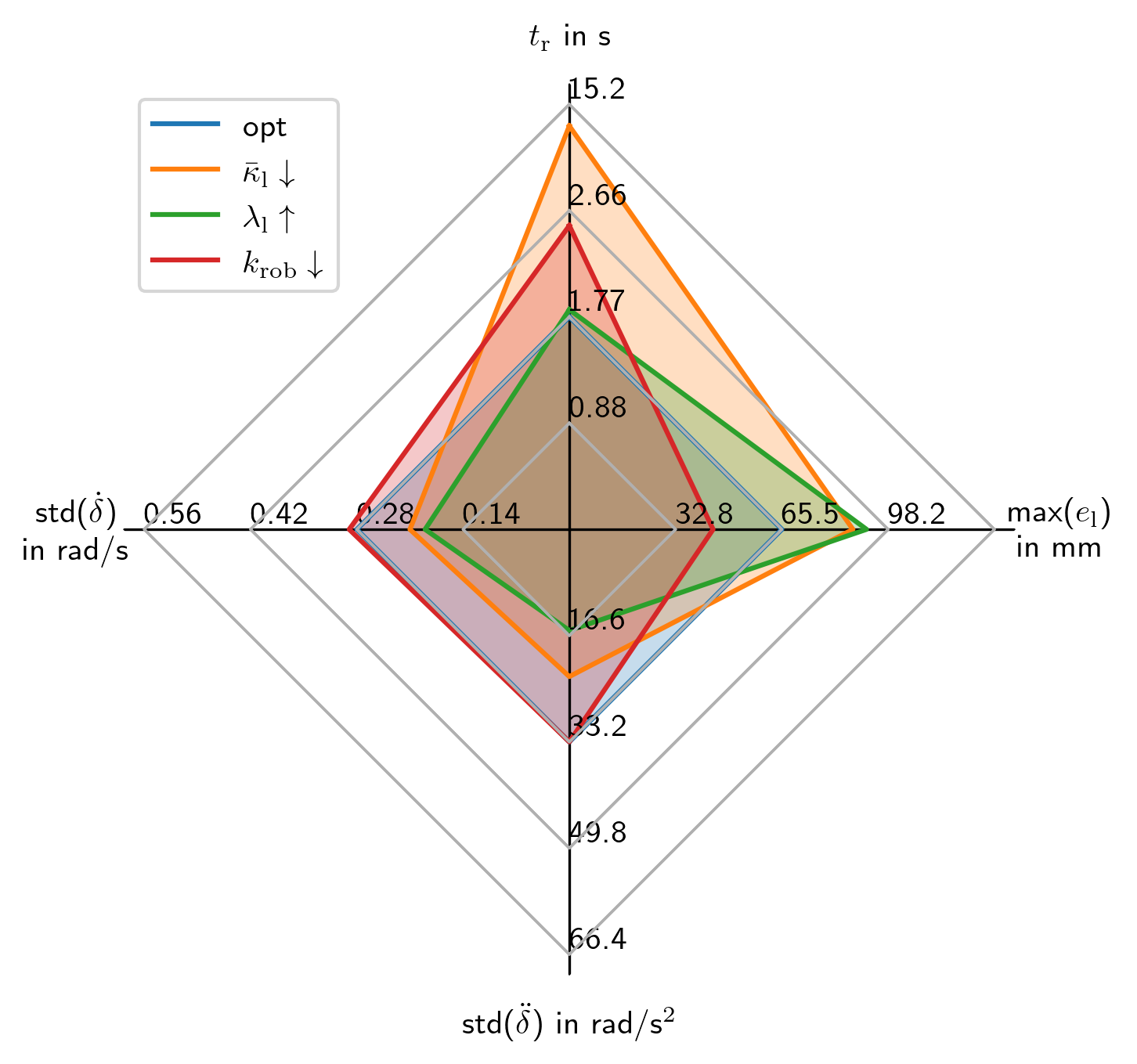}
	\caption{Performance indicators in the scenario shown in Fig. \ref{fig:simu_p} for different parameter settings.}
	\label{fig:parameter_radar}
\end{figure}

In Fig. \ref{fig:el_comp}, the control error of the lead wheel is shown for all parameter settings. While the reaching behavior differs for each parameter setting, the progression of the error after the reaching phase is quite similar.
The control performance is evaluated regarding the objectives shown in Fig. \ref{fig:trade_off}. They are represented by the following performance indicators:
\begin{itemize}
	\item The reaching time $t_\text{r}$ represents the robustness to matched disturbances. As described in \cite{robopt}, robustness to matched disturbances  is especially important for the reaching of a curved reference path. 
	\item The tracking error of the lead wheel after the reaching phase $\text{max}(e\lead)$ is representative for the suppression of noise.
	\item The maximum change of the steering angles derivative $\bar\delta''$ does also lower the values for $\delta''$ aside from the border cases. As the border case does not occur in the simulation, this objective is represented by $\text{std}(\ddot\delta)$.
	\item Similar to $\bar\delta''$, $\bar\delta'$ is represented by $\text{std}(\dot\delta)$.
\end{itemize}
The results are presented in Fig. \ref{fig:parameter_radar}.
When $\bar\kappa\lead$ is decreased
, both $\dot\delta$ and $\ddot\delta$ decrease. 
However, the sensitivity to disturbances (matched and unmatched/noise) increases which results in larger values for $t_\text{r}$ and $\text{max}(e\lead)$.\\
When $\lambda\lead$ is increased, $\ddot\delta$ decreases. While the maximum steering rate that can occur increases according to Sec.~\ref{sec:paramterization}, the actual occurring $\text{std}(\dot\delta)$ decreases. Against the expectations of a constant noise suppression as discussed in Sec.~\ref{sec:paramterization}, the control error $\text{max}(e\lead)$ increases. This is due to an increased unmatched disturbance $\psi_\text{d}$ in $e\lead' = \sin(\psi\lead + \psi_\text{d})$.\\
When $k_\text{rob}$ is decreased, $t_\text{r}$ increases as the matched disturbance $\kappa_\text{ref}$ is not fully compensated. At the same time, the noise suppression increases, which leads to a lower $\text{max}(e\lead)$.

\subsection{Comparison to other control solutions}\label{sec:benchmark}
The controller is compared to two control solutions:
\begin{enumerate}
	\item the optimal solution under constraints \labelcref{con:delta,con:ddelta}
	\item a HOSM control as presented in \cite{Incremona2017} 
\end{enumerate}
The chosen solutions satisfy constraints \labelcref{con:delta,con:ddelta}, but not constraint \labelcref{con:dddelta}. Therefore, they are compared to the controller with the $C^0$smooth output presented in this paper.\par
\subsubsection{Optimal solution}
The optimal solution is computed numerically for the given scenario. The solution is considered optimal, when the traveled path until the reference path is reached is minimal.\par

\subsubsection{HOSM control}
In \cite{Incremona2017}, a design procedure for HOSM controllers with hard constraints on the control and state variables is presented. 
This procedure is applied to design a control law for the extended system \eqref{eqn:x_ext} with constraints \labelcref{con:delta,con:ddelta}.
The first step is to define a suitable sliding manifold $\Sigma = \{\boldsymbol{x}_\text{e}: \boldsymbol{\sigma}(\boldsymbol{x}_\text{e})=0\}$ with $\boldsymbol{\sigma} = \begin{bmatrix}
	\sigma_1 & \sigma_2 & \sigma_3\end{bmatrix}$ and $\sigma_2 = \sigma_1', \sigma_3 = \sigma_2'$. The system is then transformed into the normal form by a diffeomorphism $\Omega(\boldsymbol{x}_\text{e}) = (\boldsymbol{\xi},\boldsymbol{\sigma})$:
\begin{subequations}
	\begin{align}
		\boldsymbol{\xi}' &= \phi(\boldsymbol{\xi}, \boldsymbol{\sigma})\\
		\sigma_3' &= f(\boldsymbol{\xi},\boldsymbol{\sigma})+g(\boldsymbol{\xi},\boldsymbol{\sigma})\cdot \delta'
		\end{align}
\end{subequations}
where $\boldsymbol{\xi}$ is the internal state vector.
The sliding manifold $\Sigma$ is suitable, when the following 2 assumptions hold:
\begin{enumerate}
	\item There exist upper bounds for $f(\boldsymbol{\xi},\boldsymbol{\sigma})$ and $g(\boldsymbol{\xi},\boldsymbol{\sigma})$ as well as a lower bound  for $g(\boldsymbol{\xi},\boldsymbol{\sigma})$.
	\item The constraints \labelcref{con:delta,con:ddelta} transform to
		$(\boldsymbol{\xi},\boldsymbol{\sigma}) \in \mathcal{S}(\boldsymbol{\sigma})$.
In other words, the constraint set $\mathcal{S}$ does not depend on the internal state $\boldsymbol{\xi}$.
\end{enumerate}
For the sliding manifold \eqref{eqn:control_law}, both assumptions do not hold.
Therefore, a simpler sliding manifold is chosen.
As the relative degree of the system~\eqref{eqn:x_ext} is $r=3$ (assuming $\delta'$ as input and $e$ as output), a suitable sliding manifold is not far to seek:
\begin{equation}
	\sigma_1:= e
\end{equation}
The system in normal form then is:
\begin{subequations}
	\begin{align}
		\sigma_1' &= \sigma_2\\
		\sigma_2' &= \sigma_3\\
		\begin{split}
		\sigma_3' &= -\sigma_2\frac{\sigma_3^2}{1-\sigma_2^2}+\sqrt{1-\sigma_2^2}\left(1+\lambda^2\cdot\frac{\sigma_3^2}{1-\sigma_2^2}\right)\delta'\\
		&= f(\boldsymbol{\sigma}) + g(\boldsymbol{\sigma})\cdot \delta'\end{split}
	\end{align}
\end{subequations}
where 
 $f(\sigma)$ is upper bounded ($f(\boldsymbol{\sigma})\leq F$).
For the lower bound $g(\boldsymbol{\sigma})\geq G>0$ to exist, $\sigma_2$ must be constrained. We define the constraint:
\begin{equation}
	|\sigma_2| \leq \bar\sigma_2 \Leftrightarrow |\psi| \leq \bar\psi
\end{equation}
Which results in the bounds $G$ and $F$:
\begin{subequations}
	\begin{align}
	g(\sigma(\boldsymbol{x_\text{e}})) &= |\cos\psi| \geq \cos\bar\psi = G\\
	\begin{split}
	f(\sigma(\boldsymbol{x_\text{e}})) &= -\sin\psi\cdot\left(\tan\delta/\lambda\right)^2\\ &\leq  \sin\bar\psi\cdot\left(\tan\bar\delta/\lambda\right)^2 = F
	\end{split}\end{align}
\end{subequations}

According to \cite{Incremona2017}, the resulting control law is:
\begin{subequations}
	\begin{align}
		\delta' = \begin{cases}
			-\bar\delta'\cdot\sign\left(s(\sigma) \right)\\
			-\bar\delta'\cdot\sign(\sigma_3)
		\end{cases}
	\end{align}
\end{subequations}
Where $s(\boldsymbol{\sigma})$ is the switching surface as defined in \cite{Walther2001}:
\begin{subequations}
	\begin{align}
	s(\boldsymbol{\sigma}) &= \sigma_1 + \frac{\sigma_3^3}{3\alpha^2}+\sign\left(\sigma_2+\frac{\sigma_3^2\sign(\sigma_3)}{2\alpha}\right)\\
	\alpha &= G\cdot \bar\delta' - F
	\end{align}
\end{subequations}

\begin{figure} 
	\centering
	\includegraphics[width=0.5\textwidth, trim=0cm 0.6cm 0cm 1.5cm,clip]{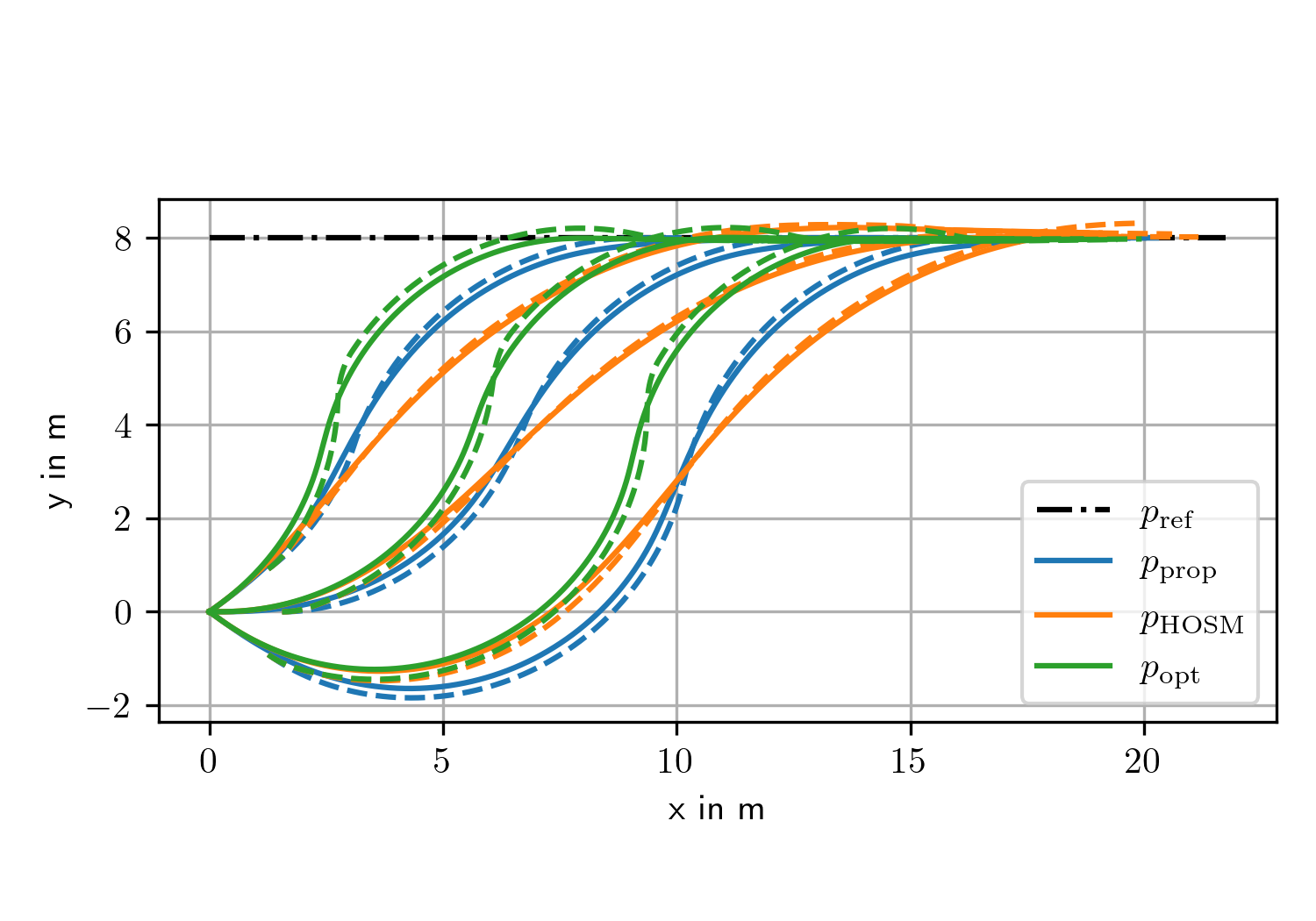}
	\caption{Reaching behavior of 3 controllers at 2 different starting positions. The solid and dashed lines depict the positions of the rear wheel and the front wheel respectively.}
	\label{fig:benchmark_e}
\end{figure}

The proposed controller as well as the optimal solution and the HOSM controller are evaluated in an ideal (undisturbed) environment for different initial tracking errors. The results are shown in Fig.~\ref{fig:benchmark_e}.\\
In the optimal solution, the front wheel overshoots the reference path such that the rear wheel reaches the reference path in shortest time.
In contrast to this, with the proposed controller the front wheel reaches the reference path faster, resulting in a slower reaching of the rear wheel.\\
For the HOSM controller, the parameter $\bar\psi$ can be used to tune between slow reaching and overshooting the reference path. The parameter is chosen optimal for start position~1 ($\psi(0)=0$), which leads to an overshooting in the other cases.\par
In Tab. \ref{Tab:roc_vx}, the traveled distance until the reference path is reached, is shown for the 3 solutions and 3 starting positions. While the traveled path of the HOSM controller is only slightly longer for starting position~1, it greatly increases in the other cases.
\begin{table}[h]
	\centering
	\begin{tabular}{l|lrrr}
			controller & $\psi(0)=$ & $0$ & $0.2\pi$ & $-0.2\pi$\\
			\hline
			optimal & & $1.42$m & $1.17$m & $1.91$m\\
			proposed & & $1.69$m & $1.38$m & $2.24$m\\
			HOSM & & $1.82$m & $2.14$m & $3.00$m\\
		\end{tabular}
	\caption{Traveled distance to reach the reference path of different controllers and for different starting positions.} 
	\label{Tab:roc_vx}
\end{table}

\subsection{Lane change maneuver}\label{sec:lanechange}
\begin{figure}  
	\centering
	\begin{subfigure}[b]{0.5\textwidth}
		\centering
		\includegraphics[width=\textwidth, trim=0cm 0cm 0.6cm 1cm,clip]{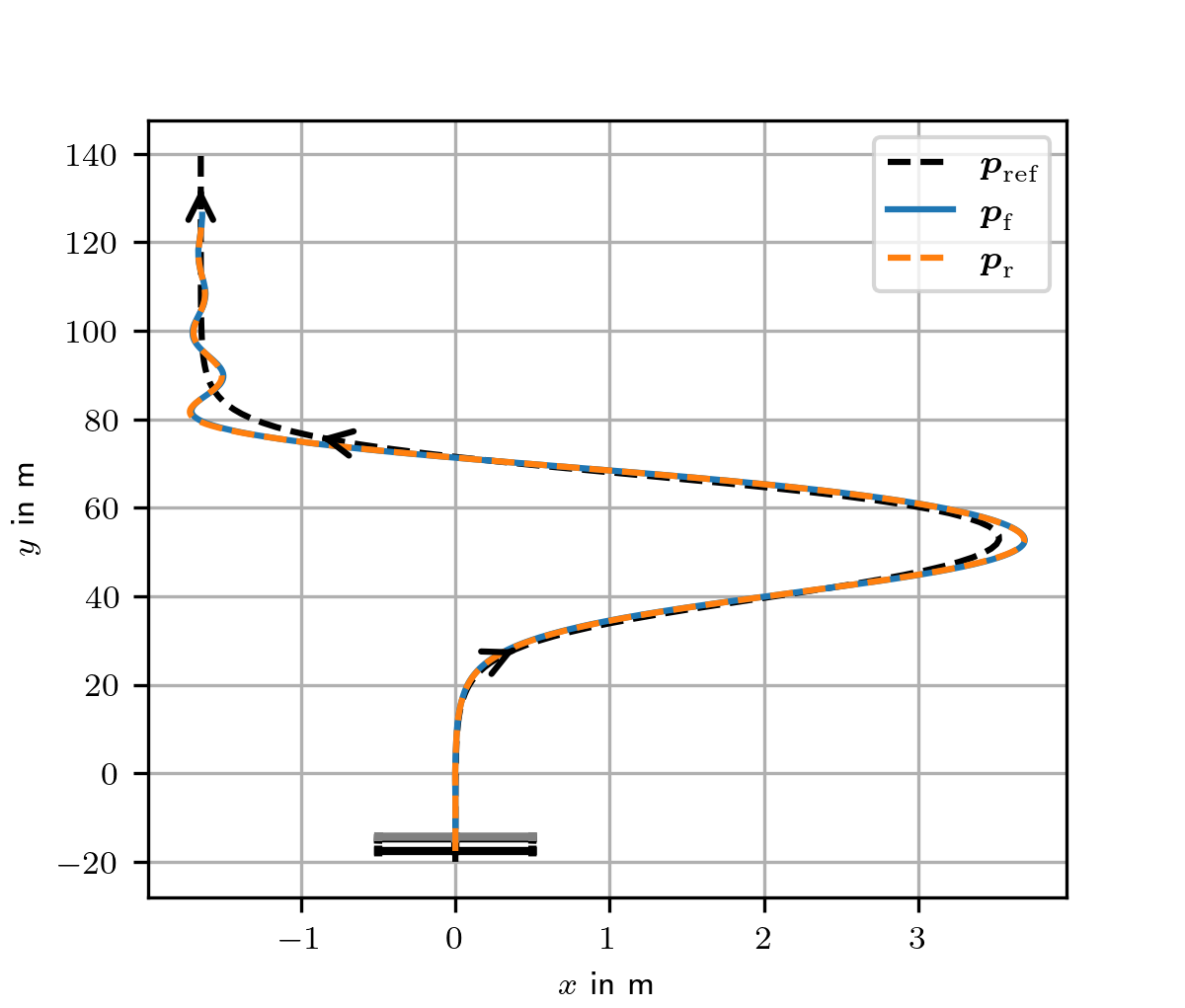}
		\caption{Reference path and path driven by the vehicle}
		\label{fig:lanechange_p}
	\end{subfigure}
	\begin{subfigure}[b]{0.5\textwidth}
		\centering
		\includegraphics[width=\textwidth, trim=0.3cm 0.2cm 0cm 0cm,clip]{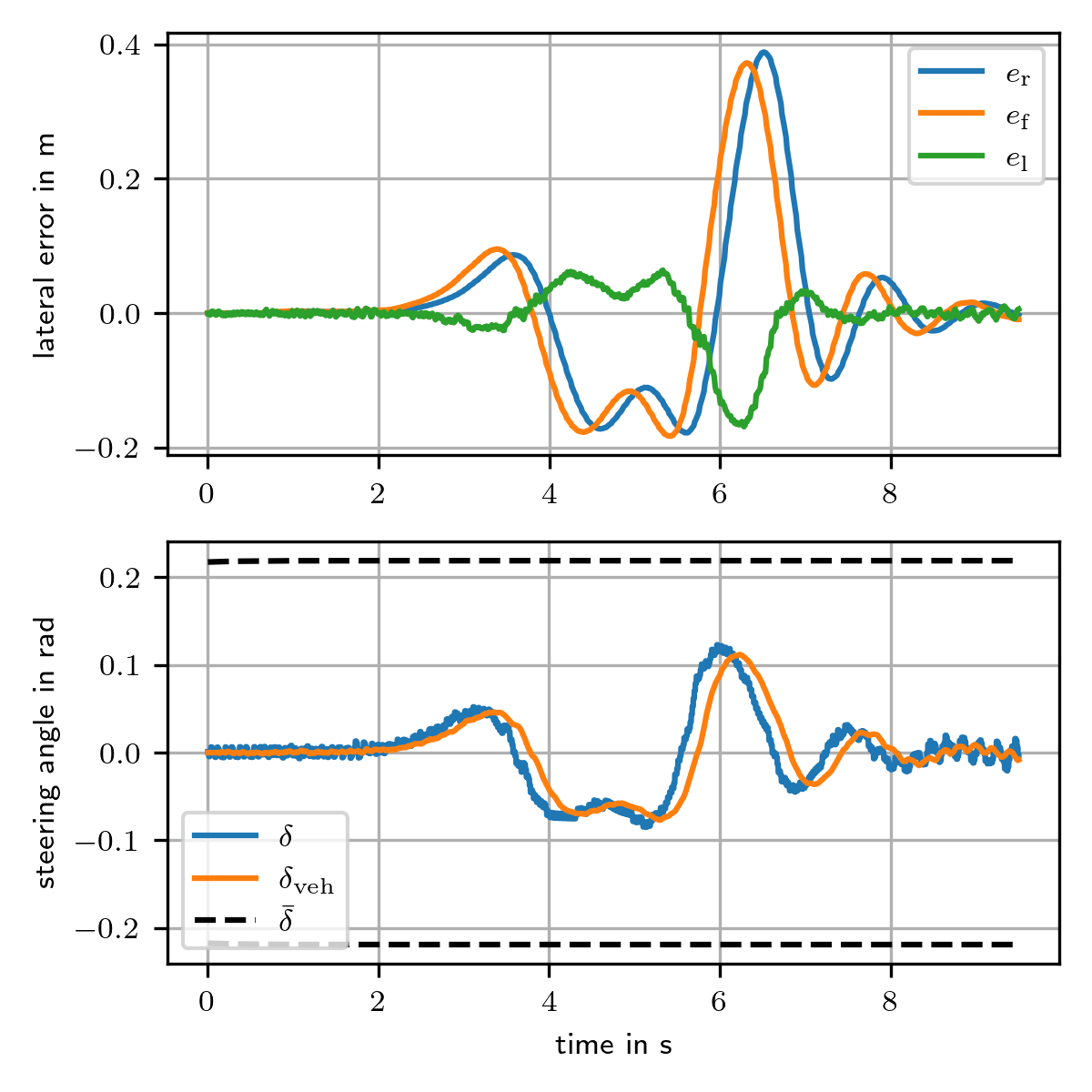}
		\caption{Tracking error and steering angle}
		\label{fig:lanechange_e}
	\end{subfigure}
	\caption{Simulation of the controller with a $C^1$ smooth output in a lane change maneuver at a velocity of $15$m/s.}
\label{fig:lanechange}
\end{figure}

The evaluation scenarios in the past two subsections where selected to show specific characteristics of the proposed controller.
To complement this, we demonstrate the controllers performance in a more common scenario, namely the lane change path defined in \cite{lanechange} and shown in Fig.~\ref{fig:lanechange_p}. The vehicle is the same kinetic single track model as used in Sec. \ref{subsec:noise}.

At a velocity of $15$m/s, the lateral slip angle reaches values of up to $0.093$rad and thus has a significant impact on the controllers performance. As can be seen in Fig.~\ref{fig:lanechange}, the vehicle overshoots the corners leading to tracking errors of up to $0.39$m. While the control error $e\lead$ returns to $0$ immediately after the curve, the lateral offset of the front and rear wheel show fading oscillations.
The behavior of parameter tuning is similar to the results of Sec.~\ref{subsec:noise}. When decreasing $\kappa\lead$ or increasing $\lambda\lead$, the steering rate and acceleration decreases while the maximum control error $e\lead$ increases. As the unmatched disturbances (in this case vehicle slip) outweigh the matched disturbance (curvature), the performance is better for smaller values for $k_\text{rob}$ (the control error decreases while the actuation effort remains unchanged).
The controller has also been evaluated for $10$m/s and $19$m/s. In these tests, the maximum control error is $0.012$m and $0.99$m respectively. The same scenario has been evaluated in \cite{newtonraphson} with a similar vehicle model.
The controller presented there performs better at $19$m/s (with a control error of $0.25$m), but worse at $10$m/s (with a  control error of $0.07$m). The reason is that this controller considers vehicle side slip directly in the controller design. It is subject to future work to incorporate estimates of the vehicle side slip for compensating this effect and thus improving the tracking performance at larger velocities.

\section{Placement in the existing design methodologies}\label{sec:sota_placment}
The presented control design approach is motivated by ideas originating from mechanical considerations (especially the system extensions discussed in Sec.~\ref{sec:dubin_ext}). Yet, the resulting design approach shows strong similarities 
to standard control design approaches. These are discussed in the following.

\subsection{Flatness based control}
For the original dynamic system \eqref{eqn:dynsys}, the lateral error $e$ is a so called flat output as defined in \cite{adamy}. We can transform the system as follows:
\begin{subequations}
	\begin{align}
		\boldsymbol{x}_\text{flat} &= \begin{bmatrix}
		e\\ e'
	\end{bmatrix} = \phi(\boldsymbol{x}) = \begin{bmatrix}
		e \\ \sin\psi
	\end{bmatrix} \\
	\boldsymbol{x}_\text{flat}' &= \begin{bmatrix}
		e' \\ \sqrt{1-e'^2}\cdot\tan\delta/\lwb
	\end{bmatrix}
	\end{align}
\end{subequations}
By designing a control law for $\delta$, we can prescribe any $C^1$
 function for $e$. 
To obtain a smooth steering angle, $e$ must be $C^2$ smooth. We therefore increase the order of the flat system to $\boldsymbol{x}_\text{flat,e} = \begin{bmatrix}
	e& e' & e''
\end{bmatrix}^T$. We get:
\begin{equation}
e'''= -e'\frac{e''^2}{1-e'^2}+\sqrt{1-e'^2}\cdot \left(1+\lambda^2\frac{e''^2}{1-e'^2}\right)\cdot\delta'
\end{equation}
The task is now to define a control law for $\delta'$ to reach $\lim_{s\to\infty} e = 0$ while satisfying the constraints \labelcref{con:delta,con:ddelta}.
To satisfy the constraint steering angle (\labelcref{con:delta}), the optimal function for $e$ is the Dubins curve as discussed in \cite{robopt}. The Dubins curve expressed in lateral distance to a reference path $e_\text{dub}$ is:
\begin{equation}\label{eqn:edub}
	e_\text{dub}'' := \sqrt{1-e_\text{dub}'^2}\cdot\bar\kappa\cdot\sign\left(\sigma(e_\text{dub}, e_\text{dub}')\right)
\end{equation}
This curve is $C^1$ smooth. To obtain a curve that is $C^2$ smooth, we increase the order of smoothness with a first order lag element:
\begin{subequations}
	\label{eqn:ddde}
	\begin{align}
		\label{eqn:dddea}
		e' &:= (e_\text{dub} - e) / \lwb\\
		\implies e''' &:= (e_\text{dub}'' - e'')/\lwb
		\label{eqn:dddec}
	\end{align}
\end{subequations}

The resulting control law then is:
\begin{equation}
		\delta' = (e'''+e'\frac{e''^2}{1-e'^2})\frac{\sqrt{1-e'^2}}{(\lambda \cdot e'')^2+(1-e'^2)} \label{eqn:ddde_ddelta}
\end{equation}
Where $e'''$ is computed from equation \eqref{eqn:ddde} and $e_\text{dub}''$ is computed from equation \eqref{eqn:edub}.\par 

By comparing \eqref{eqn:dddea} with \eqref{eqn:pt1}, it becomes clear that $e_\text{dub}$ is equal to $e\front$ defined in \eqref{eqn:efront}. By transforming the system variables $\boldsymbol{x}_\text{flat}$ back to the original system, it can be shown that the solution \eqref{eqn:ddde_ddelta} is equal to \eqref{eqn:ext1}.

\subsection{HOSM control}
The control law \eqref{eqn:cntrllaw} is a first order sliding mode control. To obtain a continuous steering angle, we applied a control law to a new virtual control $\delta'$. We will now show that the resulting control is of a form similar to the control algorithm with prescribed convergence law, as described in \cite{fridman}:
\begin{equation*}
	\delta' = \bar\delta' \cdot \sign(\zeta(\sigma_u, \sigma_u'))
\end{equation*}
Where $\sigma_u$ is the sliding variable.
The idea is to keep $\zeta(\sigma_u, \sigma_u')=0$ in the 1-sliding mode which leads to $\sigma_u\equiv 0$ reaching and remaining in the 2-sliding mode.
We define $\sigma_u$ as follows:
\begin{equation*}
	\sigma_u := e\front = e + \sin\psi\cdot\lambda
\end{equation*}
Which is twice differentiable with respect to the traveled path. For this analysis, we consider the path traveled by the front wheel $s\front$ instead of the path traveled by the rear wheel $s$.
By substituting $\sigma_u$ in \eqref{eqn:sigmaf}, we can compute $\zeta$:
\begin{equation}
	\begin{split}
		\zeta(\sigma_u, \frac{d}{ds\front}\sigma_u) =& -\sigma_u - \left(1-\sqrt{1-\left(\frac{d}{ds\front}\sigma_u\right)^2}\right)\\ & \cdot \frac{1}{(1-k_\text{rob})\cdot\bar\kappa\front}\cdot\sign\left(\frac{d}{ds\front}\sigma_u\right)
	\end{split}
\end{equation}

The 2-sliding mode $\sigma_u = d/ds\front\;\sigma_u = 0$ is reached in finite time, as was shown in \cite{robopt}. 
While $\sigma_u = e_\text{f}$ is reached in finite time, $e=0$ is only reached asymptotically.

\section{Conclusion}\label{sec:conclusion}
In \cite{robopt}, a controller based on Dubins optimal curve was presented. As this controller produces chattering on the steering angle that is not acceptable for steering a real car-like vehicle, we extended this controller to obtain a steering angle that is $C^n$ smooth and able to comply with constraints on the steering angle and its derivatives.
While the global stability under matched disturbances is maintained, the finite time reaching behavior is exchanged for asymptotic convergence.\par
This controller outputs a $C^n$ smooth steering angle and has \mbox{$n+2$} parameters to tune the control behavior regarding sensitivity to matched and unmatched disturbances as well as the actuation effort by means of the magnitude of the steering angle and its derivatives. As each parameter tunes a specific characteristic, it is easy to find the optimal parameter setting for each specific use case or environmental condition. We demonstrated this in simulation with a velocity-dependent optimal parameter setting as well as variations of this parameter setting.
Due to the implicit constraints on the steering angle and its derivatives, the controller is insensitive to actuation delays and does not tend to oscillations even in presence of systematic disturbances.
The controller is designed considering the nonlinear characteristics of the car-like vehicle, which is why global stability is achieved and the reaching behavior is close to the theoretical optimum from any starting position.\par
The control design procedure is based on kinematic considerations that are specific to the problem of controlling a car-like vehicle. The resulting control law can be interpreted as a second order sliding mode control that implicitly confines to hard inequality constraints on the steering angle and its derivatives. The resulting trajectory resembles the optimal Dubins path that is smoothed by a first order lag element.

As the control design only considers the kinematic single track model, any vehicle dynamics effects such as side slip or actuation dynamics act as disturbance on the system. The controller is suitable for scenarios with low dynamics and high demands on accuracy and guaranteed stability even in presence of large initial errors and curvy reference paths. The controller can be easily extended to be able to handle reversion maneuvers as in parking use cases.
\section{Outlook}
	While the controller remains stable in the presence of small systematic disturbances such as actuation delay or vehicle dynamics, the performance decreases at highly dynamic maneuvers. Disturbing effects such as side slip can be estimated and thus compensated in the controller, which may improve the performance of the controller significantly.
	The controller may also be extended by a steering model inversion as shown in \cite{Klauer2020}.
	
	It is also of great interest to verify, whether the controller performs similarly in a real vehicle as in the simulation.


\bibliography{IEEEabrv,journal1}

\begin{thebibliography}{10}
\providecommand{\url}[1]{#1}
\csname url@rmstyle\endcsname
\providecommand{\newblock}{\relax}
\providecommand{\bibinfo}[2]{#2}
\providecommand\BIBentrySTDinterwordspacing{\spaceskip=0pt\relax}
\providecommand\BIBentryALTinterwordstretchfactor{4}
\providecommand\BIBentryALTinterwordspacing{\spaceskip=\fontdimen2\font plus
\BIBentryALTinterwordstretchfactor\fontdimen3\font minus
  \fontdimen4\font\relax}
\providecommand\BIBforeignlanguage[2]{{%
\expandafter\ifx\csname l@#1\endcsname\relax
\typeout{** WARNING: IEEEtran.bst: No hyphenation pattern has been}%
\typeout{** loaded for the language `#1'. Using the pattern for}%
\typeout{** the default language instead.}%
\else
\language=\csname l@#1\endcsname
\fi
#2}}

\bibitem{Dominguez2016}
S.~Dominguez, A.~Ali, G.~Garcia, and P.~Martinet, ``{Comparison of lateral
  controllers for autonomous vehicle: Experimental results},'' in \emph{IEEE
  Conference on Intelligent Transportation Systems, Proceedings, ITSC}, 2016,
  pp. 1418--1423.

\bibitem{Paden2016}
B.~Paden, M.~Cap, S.~Z. Yong, D.~Yershov, and E.~Frazzoli, ``{A Survey of
  Motion Planning and Control Techniques for Self-driving Urban Vehicles},''
  \emph{IEEE Transactions on Intelligent Vehicles}, vol.~1, no.~1, pp. 33--55,
  2016.

\bibitem{Yao2020}
Q.~Yao, Y.~Tian, Q.~Wang, and S.~Wang, ``{Control Strategies on Path Tracking
  for Autonomous Vehicle: State of the Art and Future Challenges},'' \emph{IEEE
  Access}, vol.~8, pp. 161\,211--161\,222, 2020.

\bibitem{robopt}
K.~Tieber, J.~Rumetshofer, M.~Stolz, D.~Watzenig, and S.~Member, ``{Sub-optimal
  and robust path tracking: a geometric approach},'' in \emph{2021 IEEE/RSJ
  International Conference on Intelligent Robots and Systems (IROS)}, 2021, pp.
  8381--8387.

\bibitem{Cumali2019}
K.~Cumali and E.~Armagan, ``{Steering Control of a Vehicle Equipped with
  Automated Lane Centering System},'' in \emph{ELECO 2019 - 11th International
  Conference on Electrical and Electronics Engineering}.\hskip 1em plus 0.5em
  minus 0.4em\relax Institute of Electrical and Electronics Engineers Inc.,
  2019, pp. 820--824.

\bibitem{Xiaodong2019}
X.~Wu, M.~Zhang, and M.~Xu, ``{Active tracking control for steer-by-wire system
  with disturbance observer},'' \emph{IEEE Transactions on Vehicular
  Technology}, vol.~68, no.~6, pp. 5483--5493, jun 2019.

\bibitem{Loof2019}
J.~Loof, I.~Besselink, and H.~Nijmeijer, ``{Automated Lane Changing with a
  Controlled Steering-Wheel Feedback Torque for Low Lateral Acceleration
  Purposes},'' \emph{IEEE Transactions on Intelligent Vehicles}, vol.~4, no.~4,
  pp. 578--587, 2019.

\bibitem{Klauer2020}
C.~Klauer, M.~Schwabe, and H.~Mobalegh, ``{Path Tracking Control for Urban
  Autonomous Driving},'' \emph{IFAC-PapersOnLine}, vol.~53, no.~2, pp.
  15\,705--15\,712, 2020.

\bibitem{Nguyen2021}
A.~T. Nguyen, J.~Rath, T.~M. Guerra, R.~Palhares, and H.~Zhang, ``{Robust
  Set-Invariance Based Fuzzy Output Tracking Control for Vehicle Autonomous
  Driving under Uncertain Lateral Forces and Steering Constraints},''
  \emph{IEEE Transactions on Intelligent Transportation Systems}, vol.~22,
  no.~9, pp. 5849--5860, 2021.

\bibitem{Xu2020}
S.~Xu and H.~Peng, ``{Design, Analysis, and Experiments of Preview Path
  Tracking Control for Autonomous Vehicles},'' \emph{IEEE Transactions on
  Intelligent Transportation Systems}, vol.~21, no.~1, pp. 48--58, 2020.

\bibitem{Hu2020}
C.~Hu, Z.~Wang, Y.~Qin, Y.~Huang, J.~Wang, and R.~Wang, ``{Lane Keeping Control
  of Autonomous Vehicles with Prescribed Performance Considering the Rollover
  Prevention and Input Saturation},'' \emph{IEEE Transactions on Intelligent
  Transportation Systems}, vol.~21, no.~7, pp. 3091--3103, 2020.

\bibitem{conditioned_supertwist}
R.~Seeber and M.~Reichhartinger, ``{Conditioned Super-Twisting Algorithm for
  systems with saturated control action},'' \emph{Automatica}, vol. 116, p.
  108921, 2020.

\bibitem{Incremona2017}
G.~P. Incremona, M.~Rubagotti, and A.~Ferrara, ``{Sliding Mode Control of
  Constrained Nonlinear Systems},'' \emph{IEEE Transactions on Automatic
  Control}, vol.~62, no.~6, pp. 2965--2972, 2017.

\bibitem{vehicleDynamics}
R.~Rajamani, \emph{{Vehicle Dynamics and Control}}, 2nd~ed.\hskip 1em plus
  0.5em minus 0.4em\relax Springer, 2012.

\bibitem{Walther2001}
Walther, Georgiou, and Tannenbaum, ``{On the computation of switching surfaces
  in optimal control: A Gr{\"{o}}bner basis approach},'' \emph{IEEE
  Transactions on Automatic Control}, vol.~46, no.~4, p. 517, 2001.

\bibitem{lanechange}
P.~Falcone, F.~Borrelli, J.~Asgari, E.~T. Hongtei, and D.~Hrovat, ``{Predictive
  Active Steering Control for Autonomous Vehicle Systems},'' \emph{IEEE
  Transactions on Control Systems Technology}, vol.~15, no.~3, pp. 566--580,
  2007.

\bibitem{newtonraphson}
S.~Shivam, I.~Buckley, Y.~Wardi, C.~Seatzu, and M.~Egerstedt, ``{Tracking
  control by the Newton-raphson flow: Applications to autonomous vehicles},''
  in \emph{2019 18th European Control Conference, ECC 2019}.\hskip 1em plus
  0.5em minus 0.4em\relax EUCA, 2019, pp. 1562--1567.

\bibitem{adamy}
J.~Adamy, \emph{{Nichtlineare Systeme und Regelungen}}, 2nd~ed.\hskip 1em plus
  0.5em minus 0.4em\relax Berlin: Springer Vieweg, 2018.

\bibitem{fridman}
Y.~Shtessel, C.~Edwards, L.~Fridman, and A.~Levant, \emph{{Sliding Mode Control
  and Observation}}.\hskip 1em plus 0.5em minus 0.4em\relax New York: Springer,
  2014.

\end{thebibliography}
\vspace{1cm}
\begin{biography}[{\includegraphics[width=1in,height=1.25in,trim=0cm 0.5cm 0cm 1cm,clip,keepaspectratio]{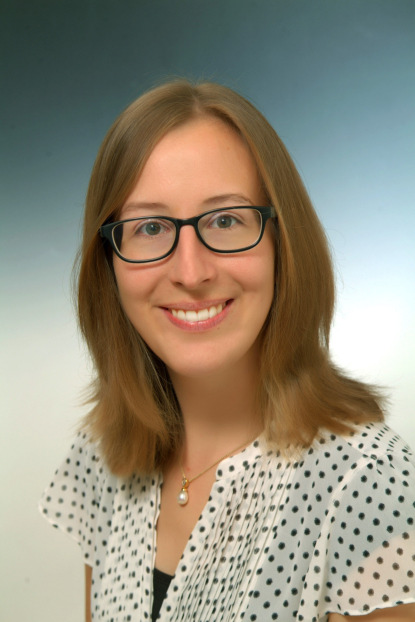}}]%
	{Karin Festl}
	received the M.Sc. degree in me-
	chanical engineering and the M.Sc. degree in information and computer engineering from the Graz University of Technology,
	Graz, Austria. She is currently working toward the Ph.D. degree in Electrical Engineering with the Graz University of Technology, Graz, Austria. She is currently part of the autonomous-systems Group with Virtual Vehicle Research GmbH. Her main research interests include automated driving, automotive control systems,
	embedded control, control-system-architecture, simulation and validation
	of automated driving.
\end{biography}
\begin{biography}[{\includegraphics[width=1in,height=1.25in,clip,keepaspectratio]{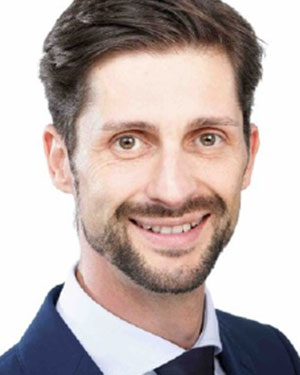}}]%
	{Michael Stolz}
	received the M.Sc. degree in me-
	chanical engineering and the Ph.D. degree in control
	engineering from the Graz University of Technology,
	Graz, Austria. He is currently a part of the Automated
	Driving Research Team with the Institute of Automa-
	tion and Control, Graz University of Technology, as
	a Postdoctoral Project Assistant. He is also a part
	of the control-systems Group with Virtual Vehicle
	Research GmbH as a key Researcher responsible
	for project management and technical guidance of
	funded project in the field of automated driving.
	He was with the automotive industry as a Development Engineer for ten
	years (one year as skill team leader in the field of automated driving). His
	main research interests include automated driving, automotive control systems,
	embedded control, control-system-architecture, algorithms for path-planning
	and control, simulation, synchronisation, optimization in control, and validation
	of automated driving.
\end{biography}
\begin{biography}[{\includegraphics[width=1in,height=1.25in,clip,keepaspectratio]{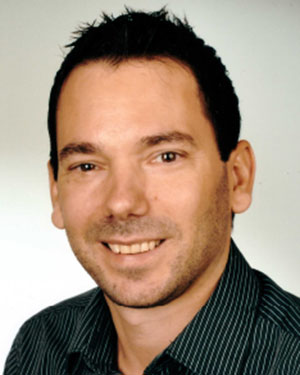}}]%
	{Daniel Watzenig}
	(Senior Member, IEEE) received
	the M.Sc. degree in electrical engineering and the
	Ph.D. degree in technical science from the Graz Uni-
	versity of Technology, Graz, Austria, in 2002 and
	2006, respectively. In 2009, he was the recipient of the
	Venia Docendi (Habilitation) for Electrical Measure-
	ment and Signal Processing. Since 2006, he has been
	the Divisional Director and Scientific Head of the
	Automotive Electronics and Software Department,
	Virtual Vehicle Research GmbH, Graz. In 2017, he
	was appointed as a Full Professor of autonomous driv-
	ing with the Institute of Automation and Control, Graz University of Technology.
	He is the Founder and a Team Leader of the Autonomous Racing Graz Team,
	one of currently six teams of the global race series Roborace. He is the Author
	or Co-Author of more than 200 peer-reviewed papers, book chapters, patents,
	and articles. His research interests include sense and control of automated
	vehicles, signal processing, multi-sensor data fusion, uncertainty estimation and
	quantification, and robust optimization.
\end{biography}

\end{document}